\documentclass[twocolumn]{aastex631}

\usepackage[T1]{fontenc}
\usepackage{microtype}

\usepackage{pifont}% http://ctan.org/pkg/pifont
\newcommand{\cmark}{\ding{51}}%

\usepackage{xspace}

\usepackage{amssymb}
\usepackage{amsmath}
\usepackage{amstext}
\usepackage{mathtools}
\usepackage{physics}

\let\tablenum\relax
\usepackage{siunitx}
\DeclareSIUnit{\parsec}{pc}
\DeclareSIUnit{\pc}{pc}
\DeclareSIUnit{\year}{yr}
\newcommand{\Msun}{\ensuremath{\mathrm{M}_{\odot}}\xspace}

\newcommand{\limepy}{\textsc{limepy}\xspace}
\newcommand{\gcfit}{\textsc{gcfit}\xspace}
\newcommand{\Nbody}{\(N\)-body\xspace}

\newcommand{\omegacen}{\(\omega\)\thinspace{Cen}\xspace}

\newcommand*\chem[1]{\ensuremath{\mathrm{#1}}}
\newcommand{\FeH}{\ensuremath{[\chem{Fe}/\chem{H}]}}

\newcommand{\BHret}{\ensuremath{\mathrm{BH}}_{\mathrm{ret}}}
\newcommand{\fbh}{\ensuremath{f_{\mathrm{BH}}}\xspace}

\newcommand{\PdotP}{\ensuremath{\dot{P}/P}\xspace}
\newcommand{\Pdot}{\ensuremath{\dot{P}}\xspace}
\newcommand{\PbdotPb}{\ensuremath{\dot{P}_{\rm b}/P_{\rm b}}\xspace}
\newcommand{\Pbdot}{\ensuremath{\dot{P}_{\rm b}}\xspace}

\newcommand{\TucAllData}{\texttt{47Tuc-AllData}\xspace}
\newcommand{\TucNoPulsars}{\texttt{47Tuc-NoPulsars}\xspace}
\newcommand{\TucNoKin}{\texttt{47Tuc-NoKin}\xspace}

\newcommand{\TerAllData}{\texttt{Ter5-AllData}\xspace}
\newcommand{\TerNoPulsars}{\texttt{Ter5-NoPulsars}\xspace}
\newcommand{\TerNoKin}{\texttt{Ter5-NoKinNoMF}\xspace}

\newcommand{\Gaia}{\textit{Gaia}\xspace}
\newcommand{\HST}{\textit{HST}\xspace}

\usepackage[normalem]{ulem}

\submitjournal{ApJ}

\shorttitle{Modeling 47 Tuc and Terzan 5 using pulsar timing}
\shortauthors{Smith et al.}
\begin{document}

\sloppy\sloppypar\raggedbottom\frenchspacing

\title{Probing populations of dark stellar remnants in the globular clusters 47 Tuc and Terzan 5 using pulsar timing}
%% you desire. In v6.31 you can include a footnote in the title.

%% Use \email to set provide email addresses. Each \email will appear on its
%% own line so you can put multiple email address in one \email call. A new
%% \correspondingauthor command is available in V6.31 to identify the
%% corresponding author of the manuscript. It is the author's responsibility
%% to make sure this name is also in the author list.
%%
%% While authors can be grouped inside the same \author and \affiliation
%% commands it is better to have a single author for each. This allows for
%% one to exploit all the new benefits and should make book-keeping easier.
%%
%% If done correctly the peer review system will be able to
%% automatically put the author and affiliation information from the manuscript
%% and save the corresponding author the trouble of entering it by hand.

\correspondingauthor{Peter J. Smith}
\email{peter.smith1@smu.ca}

\correspondingauthor{Vincent H\'{e}nault-Brunet}
\email{vincent.henault@smu.ca}

\author[0000-0002-7489-5244]{Peter J. Smith}
\affiliation{Department of Astronomy and Physics, Saint Mary's University \\
    923 Robie Street,
    Halifax, B3H 3C3, Canada}

\author[0000-0003-2927-5465]{Vincent H\'{e}nault-Brunet}
\affiliation{Department of Astronomy and Physics, Saint Mary's University \\
    923 Robie Street,
    Halifax, B3H 3C3, Canada}

\author[0000-0002-6865-2369]{Nolan Dickson}
\affiliation{Department of Astronomy and Physics, Saint Mary's University \\
    923 Robie Street,
    Halifax, B3H 3C3, Canada}

\author[0000-0002-9716-1868]{Mark Gieles}
\affiliation{Institut de Ci\`{e}ncies del Cosmos (ICCUB), Universitat de Barcelona \\
Mart\'{i} i Franqu\`{e}s 1,
08028 Barcelona, Spain}
\affiliation{ICREA,
    Pg. Lluis Companys 23,
    08010 Barcelona, Spain}

\author[0000-0002-1959-6946]{Holger Baumgardt}
\affiliation{School of Mathematics and Physics,
    The University of Queensland\\
    St Lucia, QLD 4072, Australia}

\begin{abstract}
    We present a new method to combine multimass equilibrium dynamical models and pulsar timing data
    to constrain the mass distribution and remnant populations of Milky Way globular clusters (GCs).
    We first apply this method to 47~Tuc, a cluster for which there exists an abundance of stellar
    kinematic data and which is also host to a large population of millisecond pulsars. We
    demonstrate that the pulsar timing data allow us to place strong constraints on the overall mass
    distribution and remnant populations even without fitting on stellar kinematics. Our models
    favor a small population of stellar-mass BHs in this cluster (with a total mass of $446 \substack{+75 \\ -72} \ \Msun$), arguing
    against the need for a large ($ > 2000 \ \Msun$) central intermediate-mass black hole. We then
    apply the method to Terzan~5, a heavily obscured bulge cluster which hosts the largest
    population of millisecond pulsars of any Milky Way GC and for which the collection of
    conventional stellar kinematic data is very limited. We improve existing constraints on the mass
    distribution and structural parameters of this cluster and place stringent constraints on its
    black hole content, finding an upper limit on the mass in BHs of $\sim 4000 \ \Msun$. This
    method allows us to probe the central dynamics of GCs even in the absence of stellar kinematic
    data and can be easily applied to other GCs with pulsar timing data, for which datasets will
    continue to grow with the next generation of radio telescopes.
\end{abstract}

\keywords{Black holes (162), Globular star clusters (656), Millisecond pulsars (1062), Pulsars (1306), Star clusters (1567), Stellar dynamics (1596), Stellar kinematics (1608), Stellar mass functions (1612)}

\section{Introduction} \label{sec:intro}

Pulsars have a long history of being used to investigate the mass distribution of globular clusters (GCs). Early
work from \citet{Phinney1992, Phinney} examined the effect of a pulsar's surroundings on its
measured spin (\Pdot) and orbital period derivatives (\Pbdot; for pulsars in binary systems), including quantifying the effects
of the cluster potential, Galactic potential, proper motion and intrinsic effects like magnetic
breaking. Recently, several works have presented detailed analyses of pulsar data for probing the gravitational potential of GCs (see e.g., \citealt{Prager2017a} for \object{Terzan 5}, \citealt{Freire2017} and
\citealt{Abbate2018, Abbate2019b} for \object{47 Tuc}, \citealt{Gieles2018} for NGC\,6624, \citealt{Abbate2019} for M62,\citet{Corongiu2024} for NGC 6752, and \citealt{Banares-Hernandez2024} for NGC\,5139).

Pulsars indeed present a unique opportunity to probe the central dynamics of these systems,
especially when crowding and extinction make it challenging to obtain detailed stellar kinematic
data in their central regions. Because of their extremely stable periods (spin and orbital),
measured changes in the periods of pulsars beyond their (unknown) intrinsic spin down due to magnetic breaking
can be almost entirely attributed to external factors. By performing timing measurements over long
time scales and precisely measuring the changes in their periods, we can learn about the host
potential of the pulsars. In particular, the observed period derivatives due to the changing
`Doppler shift' from the line-of-sight gravitational acceleration felt by pulsars in GCs allow
us to constrain the gravitational potential and mass distribution of GCs hosting pulsars.

Most of the studies mentioned above used single-mass dynamical models, without a mass
spectrum\footnote{With the exception of \citet{Gieles2018}, who compared the observed period
    derivatives of pulsars in NGC 6624 to predictions from multimass models but did not directly fit
    these models to the pulsar data.}. Therefore, they cannot capture the effect of mass segregation,
which is affected by the presence/absence of central black holes \citep[e.g.][]{Merritt2004,
    Mackey2008, Gill2008, Peuten2016, Peuten2017, Weatherford2018}. These studies also usually focus on
fitting models to the pulsar data, ignoring the velocity dispersion profiles, surface density
profiles and stellar mass functions which are typically used to constrain mass models of 
GCs\footnote{Although see \citet{Banares-Hernandez2024} for an application of joint modeling 
of pulsar timing data and stellar kinematics in a Jeans analysis of NGC\,5139. Their work, submitted
shortly after ours, is based on a similar methodology in many aspects, but also adopts different models
and assumptions while targeting a different cluster, and is therefore complementary.}. As an
example, \citet{Freire2017} presented a single-mass King model of 47~Tuc which was compared to the
pulsar acceleration data as well as the measured ``jerk'' (the time derivative of the acceleration) of
the pulsars. Their model does not contain an intermediate-mass black hole (IMBH) but is still able
to account for all of the pulsars' period derivatives as well as the inferred jerks for all the
pulsars within the cluster core. \citet{Corongiu2024} presented similar work for NGC 6752, using
the first and second period derivatives of five pulsars in this cluster to infer the presence of
non-luminous matter in its core and investigate the possibility of that mass residing in an IMBH.

In this work, we present new self-consistent multimass models of 47~Tuc and Terzan 5 that are fitted
both to traditional observables (velocity dispersion profiles, number density profile, local stellar
mass function observations) and to the variety of pulsar timing data available for these clusters.
The direct inclusion of the pulsar data in the likelihood function allows us to revisit and address
the previous claim from \citet{Klzlltan2017} that the pulsar accelerations favor a large central
mass in 47 Tuc in the form of an IMBH. We use updated stellar mass function data where available,
and adopt the latest constraints on the distance to our clusters from \Gaia \citep{Baumgardt2021} as
a prior on the distance parameter in our models. To properly model the effect and constrain the size
of a possible population of stellar-mass BHs, our multimass models use stellar evolution
recipes based on recent prescriptions for the masses and natal kicks of BHs, and can
therefore include realistic BH populations. While our models do not explicitly allow for an
IMBH, given that the effects of a central IMBH and a large population of centrally concentrated
stellar-mass BHs are expected to be similar beyond the sphere of influence of the IMBH,
especially when the mass fraction of the cluster in BHs is small \citep[e.g.][]{Aros2023a},
a best-fitting model without a large population of BHs would provide evidence that an IMBH is
not required to explain the dynamics of the cluster.

Since we are interested in accurately modeling the present-day mass distribution within these
clusters, equilibrium distribution function-based models are a good choice. Compared to the more
computationally expensive evolutionary models like Monte-Carlo or \Nbody models (which are limited
to relatively small grids), our multimass models offer much-increased flexibility to vary the
cluster's structural properties, stellar mass function, and population of dark remnants (including
BHs) at a small fraction of the computational cost.

Both 47 Tuc and Terzan 5  are host to large populations of pulsars. 47 Tuc is a well-studied cluster
with a great deal of stellar kinematic data, making it an ideal candidate to test our method. Terzan
5 is a heavily obscured bulge cluster which is host to the largest population of millisecond pulsars
of any Milky Way GC and which is located in the bulge, making the collection of conventional stellar
kinematic data extremely challenging. The combination of the large pulsar population and the lack of
traditional stellar kinematic data means that our method is particularly well-suited to studying the
central dynamics of this cluster. This large population of pulsars may be partially explained by the
cluster's collision rate, which is the highest of any Milky Way GC \citep[e.g.][]{Lanzoni2010}. The
central dynamics of this cluster are therefore of great interest.

The remainder of this paper is structured as follows. In Section \ref{sec:data} we describe the data
to which we fit our models. Section \ref{sec:methods} describes the models, fitting procedure and
individual likelihoods. In Section \ref{sec:results} we present our fits and discuss our results. In
Section \ref{sec:discussion} we discuss the implications of our results and compare them to other
studies. We finally summarize our findings in Section \ref{sec:conclusion}.

\section{Data}
\label{sec:data}

For this study, we use the same data that was used by \citet{Dickson2023, Dickson2024a}, with the
addition of the pulsar data described below and some additional datasets for Terzan 5. We summarize
the data below.

\subsection{Kinematics and density profiles}

\subsubsection{Proper motion dispersion profiles}

We use both {\it Hubble Space Telescope} (\HST) proper motion data and \Gaia DR3 proper motions to
constrain the kinematics of the clusters. For 47~Tuc, there are proper motion measurements from \HST
that cover the inner regions of the cluster. These data were presented in \citet{Libralato2022a}
and are split into radial and tangential dispersion profiles, allowing some leverage on the velocity
anisotropy of the cluster.

For both clusters, we also use \Gaia DR3 proper motion dispersion profiles which are based on the
membership catalogs presented in \citet{Vasiliev2021}. These profiles are split into radial and
tangential components for 47 Tuc where there is an abundance of high-probability members (the radial
and tangential profiles were derived in \citealt{Dickson2023}), but left as total proper motion
($\mu_{\rm tot}^2 = \mu_{\alpha^*}^2 + \mu_\delta^2$) for Terzan 5 where isolating high-probability
members is more difficult due to bulge contamination.

\subsubsection{Line-of-sight velocity dispersion profiles}
We use the line-of-sight velocity dispersion profiles from \citet{Baumgardt2018} to further
constrain the kinematics of the clusters. These dispersion profiles are based on archival spectra
obtained at the European Southern Observatory's (ESO) Very Large Telescope (VLT) and the Keck
observatory, supplemented with published radial velocity data from the literature from \citet{Baumgardt2017a}.

For 47 Tuc, we additionally use the line-of-sight dispersion profile presented by
\citet{Kamann2018}, who used the MUSE spectrograph \citep{Bacon2010} to collect data for 22 GCs. 

As these radial
velocity samples are dominated by bright stars, we assume that these velocity dispersion profiles
trace the kinematics of upper main-sequence and evolved stars (which we assume trace the kinematics
of giants) in our models.

\subsubsection{Number density profiles}
We use the number density profiles from \citet{DeBoer2019} and \citet{Lanzoni2010} to constrain the
size and structural parameters of the clusters. The \citet{DeBoer2019} profile is made up of a
combination of the number density profile of cluster members based on \Gaia DR2 data in the outer
regions and a surface brightness profile from \citet{Trager1995} in the central regions, which is
matched to the \Gaia number density profile in the region where the two profiles overlap.

Terzan 5 is not included in the compilation of \citet{DeBoer2019} due its position in the bulge,
where crowding and extinction are significant issues. Because of these challenges, we use the number
density profile from \citet{Lanzoni2010} which is based on a combination of data from the \HST, the
Multi-conjugate Adaptive optics Demonstrator (MAD) on VLT and the Two Micron All Sky Survey (2MASS)
which combine to cover the entire radial extent of the cluster.

The \Gaia data only includes bright stars ($m > 0.6 \ \Msun$) and the \HST and ground-based data are
likewise dominated by bright stars, so we assume that these number density profiles trace the
distribution of upper main-sequence and evolved stars in our models.

Note that there are two density profiles we could have chosen from for Terzan 5: the surface brightness
profile of \citet{Trager1995} or the number density profile of \citet{Lanzoni2010}. In their region
of overlap, these profiles do not match very well with each other, even when scaled vertically. The
\citet{Lanzoni2010} profile decreases faster in the outer regions compared to the
\citet{Trager1995} profile. We opted to use the \citet{Lanzoni2010} profile because it is based on
\HST data and modern ground-based data which should help to more reliably subtract bulge
contamination, and also because it has well-defined uncertainties. We also found that tests with the
surface brightness profile of \citet{Trager1995} resulted in best-fitting models where the surface
brightness profile and the other datasets and profiles could not be simultaneously reproduced as
well as when using the \citet{Lanzoni2010} number density profile.

\subsection{Stellar mass functions}

As a constraint on the global present-day stellar mass function of 47 Tuc, we use the
completeness-corrected stellar mass function data that was derived from archival \HST photometry by
\citet{Baumgardt2023b}. These are based on various archival \HST images 
(20 different pointings for 47 Tuc, proposal IDs shown in Figure \ref{fig:47tuc-MF}) from which
stellar number counts were derived as a function of magnitude and projected distance from the
cluster center and were then converted into stellar mass functions through isochrone fits.
For 47 Tuc, there are extensive observations which cover stars within a mass range of
\(\sim\SI{0.1}-\SI{0.8}{\Msun}\) as well as a radial range of 0-40 arcminutes from the cluster
center. The large span of radii and stellar masses allows us to constrain the varying local stellar
mass function as a function of distance from the cluster center, and therefore the degree of mass
segregation in the cluster.

The stellar mass function for Terzan 5 was derived in the same way as for 47 Tuc, but was not included
in the compilation of \citet{Baumgardt2023b} due to limitations with the data resulting from the
cluster's position in the bulge. For this cluster, we have a single mass function field, in the
infrared (filters F110W and F160W of \HST proposal 12933, PI: Ferraro) which covers a region from
0.6-1.6 arcminutes from the cluster center
and a mass range of \(\sim\SI{0.6}-\SI{0.9}{\Msun}\). While this
dataset provides weaker constraints on the model than the mass function data for 47 Tuc, it is still
useful for constraining the amount of visible mass in the cluster around its half-mass radius and
verifying that the assumed global stellar mass function in our model of Terzan 5 is reasonable.

\subsection{Pulsar data}

For 47 Tuc, we use timing
solutions from \citet{Freire2017}, \citet{Ridolfi2016} and \citet{Freire2018} which include both the
spin and orbital periods (the latter when applicable, for pulsars in binaries) and their time
derivatives. For Terzan 5, we use the timing solutions presented in \citet{Lyne2000}, \citet{Ransom2005},
\citet{Prager2017a}, \citet{Cadelano2018}, \citet{Andersen2018}, \citet{Ridolfi2021} and
\citet{Padmanabh2024a}, again including both the spin and orbital periods and their derivatives, where available.

Pulsars with non-degenerate companions are classified as either `black-widow' or `redback' systems,
where black widows have companions with masses less than $\sim 0.1 \ \Msun$ and redbacks have more
massive companions \citep{Roberts2012}. 
All redbacks and some black widows \citep[e.g.][]{Shaifullah2016} display changes in their observed orbital
periods that are likely due to Roche lobe overflow of the companion or interactions between the companion and the
pulsar wind \citep[e.g.][]{Thongmeearkom2024a}.
This orbital variability can also affect the measured spin period derivatives, and is harder to correct for in redbacks
where the orbital variability is ubiquitous and complex.
The observed changes in the spin periods of redbacks could therefore be
incorrectly interpreted as effects from the cluster potential,
so we follow \citet{Prager2017a} and exclude redback systems from our analysis. Among the well-timed
pulsars, this means we exclude pulsar \emph{W} from 47 Tuc and pulsars \emph{A}, \emph{P}, \emph{ad}
and \emph{ar} from Terzan 5. The pulsar data is summarized in Tables \ref{tab:pulsars_47tuc} and
\ref{tab:pulsars_ter5} in the Appendix.

Finally, we make use of the Australia Telescope National Facility's pulsar
database\footnote{\url{http://www.atnf.csiro.au/research/pulsar/psrcat}} presented by
\citet{Manchester2005} in order to build a representative population of Galactic MSPs which we use
to estimate the probability distribution for the intrinsic spin-down of the pulsars as a function of
their spin period (see Section~\ref{sec:pulsarlikelihoods}).

\section{Methods}
\label{sec:methods}

\subsection{Models}

To model the dynamics and mass distribution of 47~Tuc and Terzan~5, we use the
\gcfit\footnote{\url{https://github.com/nmdickson/GCfit/}} package, recently presented by
\citet{Dickson2023, Dickson2024a}. This package couples  a fast mass evolution algorithm with
the \limepy\footnote{\url{https://github.com/mgieles/limepy/}} family of models presented by
\citet{Gieles2015}. We refer readers to these papers for a detailed description of the models, and
provide a brief summary here.
The \limepy models are a set of distribution function-based equilibrium models that are isothermal
for the most bound stars near the cluster center and described by polytropes in the outer regions
near the escape energy. The models have been extensively tested against \Nbody models
\citep{Zocchi2016, Peuten2017} and their multimass version is able to effectively reproduce the
effects of mass segregation. Their suitability for mass modeling of GCs has been tested
on mock data \citep{Henault-Brunet2019}, and they have recently been applied to real datasets as
well \citep[for example,][]{Gieles2018, Henault-Brunet2020, Dickson2023, Dickson2024a}.

The input parameters needed to compute our models include the dimensionless central potential
$\phi_0$, the truncation parameter $g$\footnote{Several well-known classes of models are reproduced
    by specific values of $g$: Woolley models \citep{Woolley1954a} have $g=0$, King models
    \citep{King1966a} $g=1$, and (non-rotating) Wilson models \citep{Wilson1975a} $g=2$.}, the anisotropy radius $r_a$
which determines the degree of radial anisotropy in the models, $\delta$ which sets the mass
dependence of the velocity scale and thus governs the degree of mass segregation, and finally the
specific mass bins to use as defined by the mean stellar mass ($m_j$) and total mass ($M_j$) of each
bin, which together specify the stellar mass function. In order to scale the model units to physical
units, the total mass of the cluster $M$ and a size scale (the half-mass radius of the cluster
$r_{\rm h}$) are provided as well. Finally, we provide the distance to the cluster ($D$) which is used in
converting between angular and linear quantities.

In order to generate the input mass bins (the $m_j$ and $M_j$ sets of input values) for the
multimass \limepy models, the \gcfit models use the \texttt{evolve{\_}mf} algorithm, originally
presented by \citet{Balbinot2018} and updated in \citet{Dickson2023}. This algorithm combines
precomputed grids of stellar evolution models, isochrones and initial-final mass relations to model
the evolution of a given initial mass function (IMF), including the effects of stellar evolution as well
as (optionally) mass loss due to escaping stars and dynamical ejections. The algorithm returns a
binned mass function at a requested evolutionary time, for specified metallicity, ideal for use as an input in the \limepy
models.

We parameterize the mass function as a three-segment broken power law with break points at $0.5 \ \Msun$ and $1.0
    \ \Msun$. We provide to \texttt{evolve{\_}mf} the IMF slopes\footnote{Throughout this work we adopt the convention that $\xi(m){\equiv{\rm d}N/{\rm d}m} \propto m^{-\alpha}$ such that a positive value of $\alpha$ gives a decreasing power-law slope.} ($\alpha_1$,
$\alpha_2$ and $\alpha_3$) and break points, the cluster age, metallicity and initial escape
velocity.
We adopt the same methodology as \citet{Dickson2023} to determine the initial
escape velocities of our clusters. Briefly, we run an initial fit with an initial guess of the
escape velocity and use the present-day escape velocity of this preliminary fit to set the
initial escape velocity of the cluster. We use double the present-day value as our estimate for
the initial escape velocity, which accounts for adiabatic expansion of the cluster after
mass-loss due to stellar evolution. We note that after the initial fit, changing the initial escape velocity by
$20 \ \mathrm{km \ s}^{-1}$ in either direction has no discernible effect on the final model.
We additionally specify parameters which control the mass loss (if any)
due to escaping stars and the specific binning to be used when returning the final discrete
mass-function bins. We finally provide the black hole retention fraction ($\BHret$) which
controls the percentage of the mass in black holes initially created from the initial mass
function that is retained to the present day after natal kicks and dynamical ejections. 
We first eject primarily low-mass BHs through natal kicks and then eject the rest of
the required mass by ejecting the most massive BHs first, capturing the effect of dynamical
ejections (see \citealt{Dickson2023} for details). For this study we
do not model the mass loss due to escaping stars, so we set this mass loss due to escaping stars
to be zero, and we are effectively specifying the present-day mass function for low-mass stars,
not their IMF.

\subsection{Fitting}

The \gcfit package provides a uniform interface for fitting the coupled \limepy and
\texttt{evolve{\_}mf} models to a variety of observables using either a Markov Chain Monte Carlo
(MCMC) or nested sampling algorithm. For this work, we use the nested sampling algorithm, which is
implemented using the \texttt{dynesty} package \citep{Speagle2020,Koposov2023}. For the majority of
our parameters we adopt wide, uniform priors, with the exception of the distance where we adopt the
measurements of \citet{Baumgardt2021} as Gaussian priors and the mass function power-law slopes where we adopt the physically motivated priors described in \citet{Dickson2023}. We list the priors in Table
\ref{tab:priors}.

\begingroup
\begin{table}
    \centering
    \caption{Model parameters and their priors. Most priors are uniform and are chosen to bound the
        parameters around a reasonable range of values for both clusters. For the mass function
        slopes, we add the additional constraint that $\alpha_2$ must be steeper than $\alpha_1$
        and $\alpha_3$ steeper than $\alpha_2$. For the distance, we use a Gaussian prior with the
        distance measurement from \citet{Baumgardt2021} and its uncertainty providing the mean
        $\mu$ and standard deviation $\sigma$ of the prior. These distances are $D=4.521 \pm 0.031$ \si{kpc} and $D=6.62 \pm 0.15$ \si{kpc} for 47~Tuc and Terzan~5 respectively.
    \label{tab:priors}}
    \begin{tabular}{l l l}

        \hline
        Parameter                                & Prior Form & Value              \\
        \hline
        $\phi_0$                                 & Uniform    & [0.1, 15]          \\
        $M [10^6 \ \Msun] $                      & Uniform    & [0.01, 3]          \\
        $r_h$ [pc]                               & Uniform    & [0.5, 15]          \\
        $\log_{10}{\left(r_a / {\rm pc}\right)}$ & Uniform    & [0, 8]             \\
        $g$                                      & Uniform    & [0, 3.5]           \\
        $\delta$                                 & Uniform    & [0.3, 0.5]         \\
        $s^2$                                    & Uniform    & [0, 20]            \\
        $F$                                      & Uniform    & [1, 7.5]           \\
        $\alpha_1$                               & Uniform    & [-1, 2.35]         \\
        $\alpha_2$                               & Uniform    & [-1, 2.35] and $\ge \alpha_1$  \\
        $\alpha_3$                               & Uniform    & [1.6, 4] and $\ge \alpha_2$    \\
        $\BHret [\%]$                            & Uniform    & [0, 100]           \\
        $D [\mathrm{kpc}] $                      & Gaussian   & BV21               \\
        \hline
    \end{tabular}
\end{table}
\endgroup

\subsection{Likelihoods}

The majority of the likelihood functions we use for different datasets are Gaussian likelihoods.
Provided below is the log-likelihood for velocity dispersion profile data as an example, but all
other likelihoods are of a similar form\footnote{We note that in \citet{Dickson2023} the leading minus sign in the log-likelihood functions was missing from the text.}:

\begin{equation}
    \ln \left(\mathcal{L}\right)=-\frac{1}{2}
    \sum_{i=1}^{N_{\rm p}}\left\{\frac{\left[\sigma_{\mathrm{obs}}(r_i)
            -\sigma_{\mathrm{model}}(r_i)\right]^{2}}{\delta \sigma_{\mathrm{obs}}^{2}(r_i)}
    -\ln \left[\delta \sigma_{\mathrm{obs}}^{2}(r_i)\right]\right\},
    \label{eq:likelihood}
\end{equation}
where $\mathcal{L}$ is the likelihood, $N_{\rm p}$ is the number of data points, $\sigma_{\mathrm{obs}}$
is the measured velocity dispersion, $\sigma_{\mathrm{model}}$ is the model velocity dispersion at the corresponding radius, $r$ is the projected distance from
the cluster center, and $\delta \sigma_{\mathrm{obs}}$ is the uncertainty in the velocity
dispersion. The likelihoods for other observables are formulated in the same way, and the specifics
are discussed in \citet{Dickson2023} as well as in the \gcfit
documentation\footnote{\url{gcfit.readthedocs.io}}. The total log-likelihood is the sum of all the
log-likelihoods for each set of observations.

For the mass function and number density profile likelihoods, we include additional nuisance
parameters and scaling terms. For the number density data, we introduce a parameter $s^2$ which is
added in quadrature to the existing measurement uncertainties. This parameter allows us to add a
constant contribution to all values in the dataset, effectively lowering the weight of the data located farthest from the cluster center where the number density is
lowest. This allows us to account for limitations in the models such as the effects of potential
escapers near the cluster tidal boundary that the \limepy models do not account for (see
\citealt{Claydon2019} for a discussion of potential escapers in equilibrium models).

Finally, for the number density profile data, we make an additional modification to equation~
(\ref{eq:likelihood}) and introduce a scaling factor $K$ which allows us to fit only on the shape of
the number density profile instead of the absolute values. $K$ is defined as follows (see section
3.3 of \citealt{Henault-Brunet2020} for a complete explanation) and the model data points
($\Sigma_{\mathrm{model},i}$) are multiplied by this value before they are compared to the data:

\begin{equation}
    K= \frac{\sum_{i=1}^{N_{\rm p}} \Sigma(r)_{\mathrm{obs}, i} \Sigma(r)_{\mathrm{model}, i} / \delta
    \Sigma(r)_{\mathrm{model}, i}^{2}} {\sum_{i=1}^{N_{\rm p}}\left(\Sigma(r)_{\mathrm{model}, i}\right)^{2} / \delta \Sigma(r)_{\mathrm{obs}, i}^{2}},
    \label{eq:scaling}
\end{equation}
where $N_{\rm p}$ is the number of data points, $\Sigma(r)_{\mathrm{obs}, i}$ are the number density
measurements, $\Sigma(r)_{\mathrm{model}, i}$ are the model number densities at the corresponding radii
and $\delta \Sigma(r)_{\mathrm{obs},i}$ are the uncertainties on the number density measurements where $r$ is the projected radius of a given measurement of $\Sigma(r)_{\mathrm{obs}, i}$.

We discuss the likelihoods for the pulsar timing data and the stellar mass function data separately
in the subsections below.

\subsubsection{Pulsars}
\label{sec:pulsarlikelihoods}

Pulsar period derivatives, as measured by an observer, are made up of several distinct components,
with contributions from the cluster's gravitational potential, the gravitational potential of the
Milky Way, the pulsar's proper motion, intrinsic effects like magnetic breaking, and the changing
dispersion measure between the pulsar and the observer. The effects of the cluster's proper motion
and the Galactic potential are fairly well constrained based on the pulsar's position and motion in
the Galaxy but, the effects of processes like magnetic breaking which are intrinsic to the pulsar
itself require more careful consideration. The breakdown of
the measured period derivative $(\PdotP)_{\rm obs}$ into separate components is:
\begin{equation}
    \left(\frac{\dot{P}}{P}\right)_{\rm obs} = \left(\frac{\dot{P}}{P}\right)_{\rm int} + \frac{a_{{\rm cl}, z}}{c} +
    \frac{a_{{\rm G}}}{c} + \frac{a_{\rm S}}{c} + \frac{a_{\mathrm{DM}}}{c},
    \label{eq:pular-components}
\end{equation}
where $(\PdotP)_{\rm int}$ is any change in period due to the effects intrinsic to the pulsar like
magnetic breaking, $c$ is the speed of light, $a_{{\rm cl}, z}$ the line-of-sight acceleration of the pulsar due
to the cluster's gravitational potential and is the quantity we are most interested in, $a_{\rm G}$ is the
acceleration of the pulsar along the line of sight due to the Galaxy's gravitational potential,
$a_{\rm S}$
is the `Shklovskii' effect \citep{Shklovskii1970}, an apparent acceleration due to the proper motion of the pulsar and $a_{\mathrm{DM}}$ is
the effect of the changing dispersion measure between the pulsar and the observer.

For each of the components in equation~(\ref{eq:pular-components}), we explain below how we calculate
either a point estimate or a probability distribution for the quantities of interest, and how we
combine these to obtain a probability distribution for the measured period derivative of a pulsar,
given a model (i.e. the likelihood).

\begin{figure}
    \centering
    \includegraphics[width=0.8\linewidth]{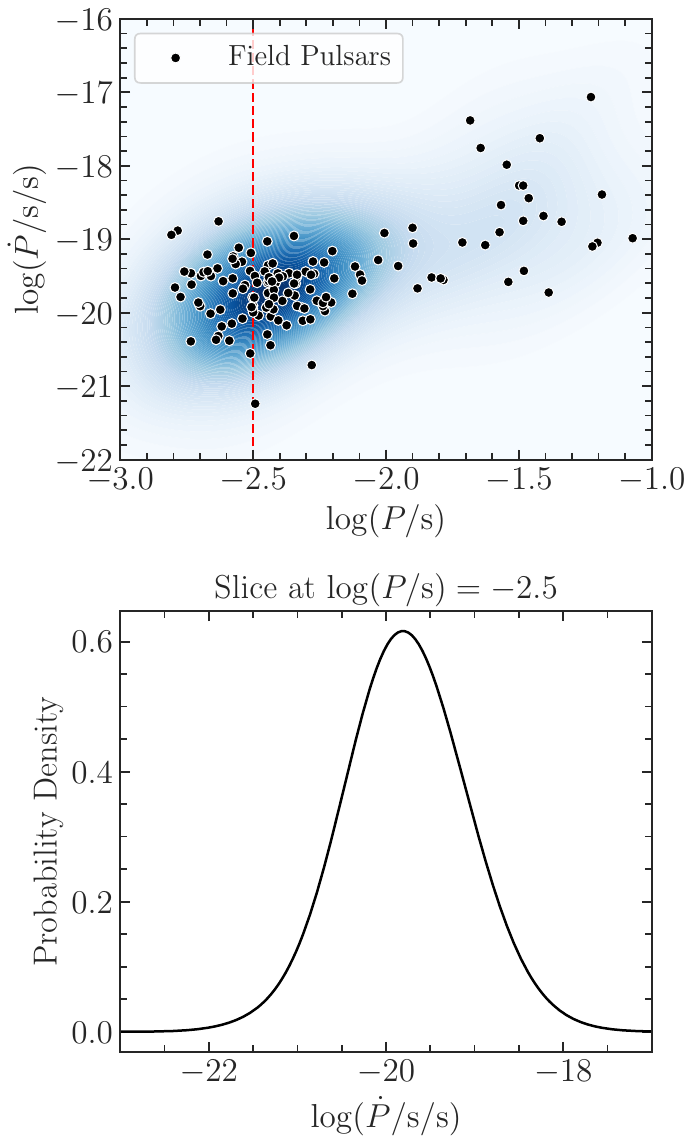}
    \caption{{\it Top}: The $P$-$\dot{P}$ plane for the field pulsars with the Gaussian KDE of the
        $P$-$\dot{P}$ distribution shown in blue. {\it Bottom}: an example of a slice from this KDE.
        The slice is taken at $\log{P}=-2.5$ s and shows the distribution of $\log{\dot{P}}$ values
        for pulsars with this period.
    \label{fig:field_pulsars}}
\end{figure}

All pulsars have some intrinsic spin-up or spin-down caused by processes like magnetic breaking or
active accretion. We exclude any redback pulsars (pulsars with massive, non-degenerate companions)
from this work and therefore  assume that none of the pulsars are actively accreting and that any
intrinsic effects are purely in the spin-down direction (positive  $(\dot{P}/P)_{\rm int}$ term in
Equation \ref{eq:pular-components}). To estimate the probability distribution for the intrinsic
spin-down distribution, we assume that the intrinsic spin-down of cluster pulsars follows the same
distribution as the Galactic field pulsars, and that it is dependent only on their period. We use
the ATNF pulsar catalog\footnote{\url{https://www.atnf.csiro.au/research/pulsar/psrcat/}}
\citep{Manchester2005} to build a distribution of possible $\dot{P}$ values for a given value of $P$
using the Galactic field pulsars as a reference (for which the period derivative can be directly
linked to the intrinsic spin-down after correcting for Galactic and proper motion contributions due
to there being no cluster acceleration for these pulsars). We compute a Gaussian kernel density
estimator (KDE) in the field $P$-$\dot{P}$ space, which is sliced along each cluster pulsar's period
to extract a distribution of intrinsic values. We show the field pulsars in the $P$-$\dot{P}$ plane,
the KDE and an example of a slice from this KDE in Figure \ref{fig:field_pulsars}.

The next two components, $a_{{\rm cl}, z}/c$ and $a_{\rm G}/c$, are fundamentally similar in that they are both
manifestations of the Doppler effect. In the typical case, we infer a star's radial velocity by
measuring the frequency shift of some known spectral feature, but in the case of pulsars, we instead
measure the acceleration of the pulsar along the line of sight by measuring the rate of change of
the pulsar's period.

We will first look at $a_{\rm G}/c$, the effect of the Galaxy's gravitational potential on the pulsar's
period derivative. The acceleration due to the Galaxy's potential is a function of the pulsar's
position in the Galaxy and the Galaxy's mass distribution. We use the \textsc{Gala} package
\citep{Price-Whelan2017} to calculate the acceleration due to the Galactic potential at each
cluster's position. We adopt the 
\textsc{MilkyWayPotential2022} potential from \textsc{Gala} as well as the cluster positions measured by \citet{Vasiliev2021} who used
\emph{Gaia} EDR3 data to measure the positions (including distances) and kinematics of Milky Way
GCs. After projecting this acceleration along the line of sight, the effect on the
period derivative is the following:
\begin{equation}
    \dot{P}_{\rm G} = \frac{a_{\rm G} P}{c},
\end{equation}
where $\dot{P}_{\rm G}$ is the contribution to the period derivative due to the Galactic potential.

The effect of the cluster's gravitational potential on the pulsar's period derivative (the $a_{{\rm cl}, z}/c$
term in equation~\ref{eq:pular-components}) is the effect we are most interested in as it helps us
to constrain the internal mass distribution of the cluster. For a pulsar with a known 3-D position
within the cluster, the period derivative due to the cluster's gravitational potential
($\dot{P}_{\rm cl}$) will simply be:

\begin{equation}
    \dot{P}_{\rm cl} = \frac{a_{{\rm cl},z} P}{c}.
\end{equation}
Because the pulsar  position along the line of sight is unknown, we need to generate a probability distribution over possible line-of-sight
accelerations for a given pulsar at a given projected radius for a given model, based on the
enclosed mass over the full range of possible line-of-sight positions. We can then weight this
distribution by the probability of a pulsar being at a given position $z$ along the line of sight to generate a
probability distribution for $\dot{P}_{\rm cl}$ given a projected radius for a given model.

In order to generate the probability distribution of a pulsar being at a given position along the line of sight we insert a tracer mass bin into the \limepy models. This tracer mass bin is a single mass
bin with a mean mass of $1.6 \  \Msun$\footnote{ We use $1.6 \ \Msun$ as the tracer mass because
    most of the pulsars (which we assume to have masses of $1.4 \ \Msun$) have binary companions with typical masses of $0.2 \ \Msun$ (see e.g.
    \citealt{Freire2017}).} and a negligible total mass. This allows us to calculate the (line-of-sight) density profile of
pulsar-mass objects in our model at a given projected radius. This profile is proportional to the probability that a pulsar has a given line-of-sight acceleration, given the model parameters, which indirectly provides constraints on the position of the pulsar along the line-of-sight.
Combining the line-of-sight density profile with the line-of-sight acceleration profile we then obtain a
probability distribution over the range of period derivatives for a given projected radius for a
given model.

\begin{figure*}
    \centering
    \includegraphics[width=1.0\textwidth]{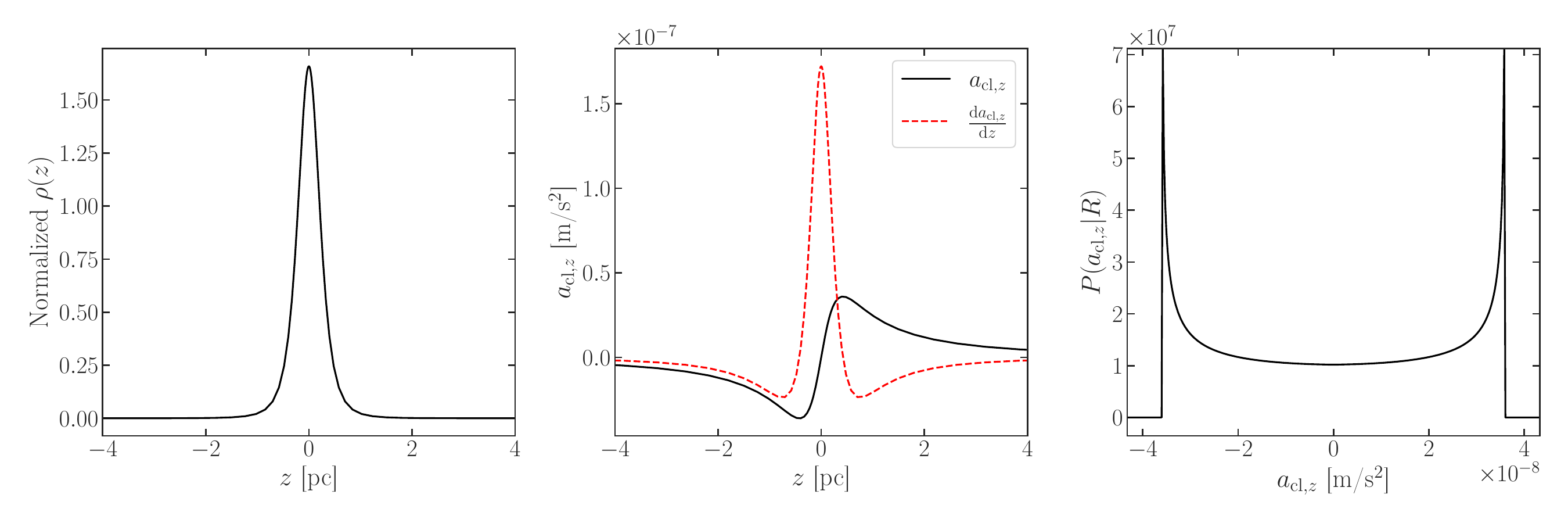}
    \caption{An example of the combination of the  line-of-sight density and acceleration profiles
        leading to the probability distribution of line-of-sight accelerations. \emph{Left panel:}
        The normalized line-of-sight density profile for pulsar-mass objects for a model fit to Terzan 5 at the
        projected radius of pulsar \emph{aa}. \emph{Middle panel:} The
        line-of-sight acceleration profile for the same model and projected radius. The derivative
        of this profile, used in the calculation of the probability distribution of line-of-sight
        accelerations is shown with a red, dashed line. \emph{Right panel:} The probability
        distribution for the line-of-sight acceleration for the same model, at the same projected
        radius. This distribution is a  result of combining the density profile with the derivative
        of the acceleration profile.
    \label{fig:ac_distribution}}
\end{figure*}

We show the expression from which we can calculate this probability distribution in
equation~(\ref{eq:paz}):

\begin{equation}
    P\left(a_{{\rm cl}, z} \mid R_i\right) \propto \frac{\mathrm{d} m}{\mathrm{d} a_{{\rm cl},z}}=\frac{\mathrm{d} m}{\mathrm{d} z} \abs{\frac{\mathrm{d} z}{\mathrm{d} a_{{\rm cl}, z}}}=\frac{\rho(z)}{\left|{{\rm d} a_{{\rm cl},z}/{\rm d} z}\right|},
    \label{eq:paz}
\end{equation}

where $P\left(a_{{\rm cl},z} \mid R_i\right)$ is the probability of a given line-of-sight acceleration
measurement ($a_{{\rm cl},z}$) for a projected radius $R_i$, $m$ is mass column density of
pulsar-mass objects along the line of sight at projected radius $R_i$, $a_{{\rm cl},z}$ is the line-of-sight
acceleration for a given line-of-sight position and $\rho(z)$ is the mass density of pulsar-mass
objects at a given line-of-sight position. We note that, as stated on the right-hand side of Equation \ref{eq:paz}, this
probability distribution is not normalized. After constructing this distribution, we explicitly normalize
it such that it behaves like a probability density function. Each of the  quantities on the right-hand side of
Equation~(\ref{eq:paz}) ($\rho(z)$, ${\rm d} a_{{\rm cl},z}/{\rm d}z$) are calculated for each \limepy
model. We show the combination of these distributions in Figure \ref{fig:ac_distribution}.

The Shklovskii
effect ($a_{\rm S}/c$) is the effect of the pulsar's proper motion on its observed period derivative. 
Any transverse motion of a pulsar acts to increase the distance to the pulsar,
regardless of the direction of motion. A constant transverse motion results in a non-linear 
increase in the distance, manifesting as an apparent line-of-sight acceleration 
\citep[e.g.][]{Verbiest2008}. This effect is calculated as:
\begin{equation}
    \dot{P}_{\rm S} = \frac{a_{\rm S}P}{c} = \frac{\mu^2 D P}{c},
\end{equation}
where $\dot{P}_{\rm S}$ is the rate of change of the period due to the Shklovskii effect and $\mu$
is the proper motion. We use the cluster's bulk proper motion  to calculate this effect and again
adopt the measurements of \citet{Vasiliev2021}\footnote{$\mu_{\alpha^*} = 5.253 \pm 0.008 \
        \mathrm{mas \ yr^{-1}}$ and $\mu_\delta = -2.557 \pm 0.008 \	\mathrm{mas \ yr^{-1}}$ for 47 Tuc and
    $\mu_{\alpha^*} = -1.864 \pm 0.030  \ \mathrm{mas \ yr^{-1}}$ and $\mu_\delta = -5.108 \pm 0.027 \
        \mathrm{mas \ yr^{-1}}$ for Terzan 5.}. Finally, $D$ is the distance to the cluster, one of the parameters
which we allow to vary in our fitting. This effect is of order $a_{\rm S} \sim 10^{-11} - 10^{-10} \
    \mathrm{m \ s^{-2}}$ which is negligible compared the acceleration due to the cluster
potential which is of order
$10^{-9} - 10^{-8} \ \mathrm{m \ s^{-2}}$ (e.g. Figure \ref{fig:ac_distribution}).

The final component, $a_{\rm DM}/c$, is the effect of the changing dispersion measure
between a pulsar and the observer. The total amount and distribution of the ionized gas along our
line of sight is not necessarily constant over the full time-span over which these observations were
performed and small changes in the total dispersion measure between us and the cluster can cause
small variations in the observed period derivatives (see, for example, \citealt{Prager2017a}).
This effect is stochastic, meaning it is
unlikely to bias the timing solution in one direction or the other. Furthermore, the magnitude of
this effect is expected to be very small, on the order of $10^{-13} \ \mathrm{m\ s^{-2}}$
\citep{Prager2017a}, several orders of magnitude smaller than the typical acceleration from the cluster potential, therefore we do not consider it in our analysis.

One potential contribution to the observed values of \PdotP that we do not model is the acceleration
and its higher-order derivatives caused by nearby stars in the dense core of the cluster.
For the acceleration of the pulsars in particular, this effect has been shown to be typically
$\sim 2$ orders of magnitude smaller than the mean-field acceleration from the cluster potential
as a whole \citep{Phinney, Prager2017a}. This effect is however relevant for the higher order
derivatives, where nearby stars contribute at a similar level to the bulk cluster potential
\citep{Blandford1987, Gieles2018}. It is for this reason that while higher-order period derivatives
are measured for many of the pulsars we use in this work, we chose not to consider these measurements
in our determination of the mass distributions of our clusters.

\begin{figure*}
    \centering
    \includegraphics[width=1.0\linewidth]{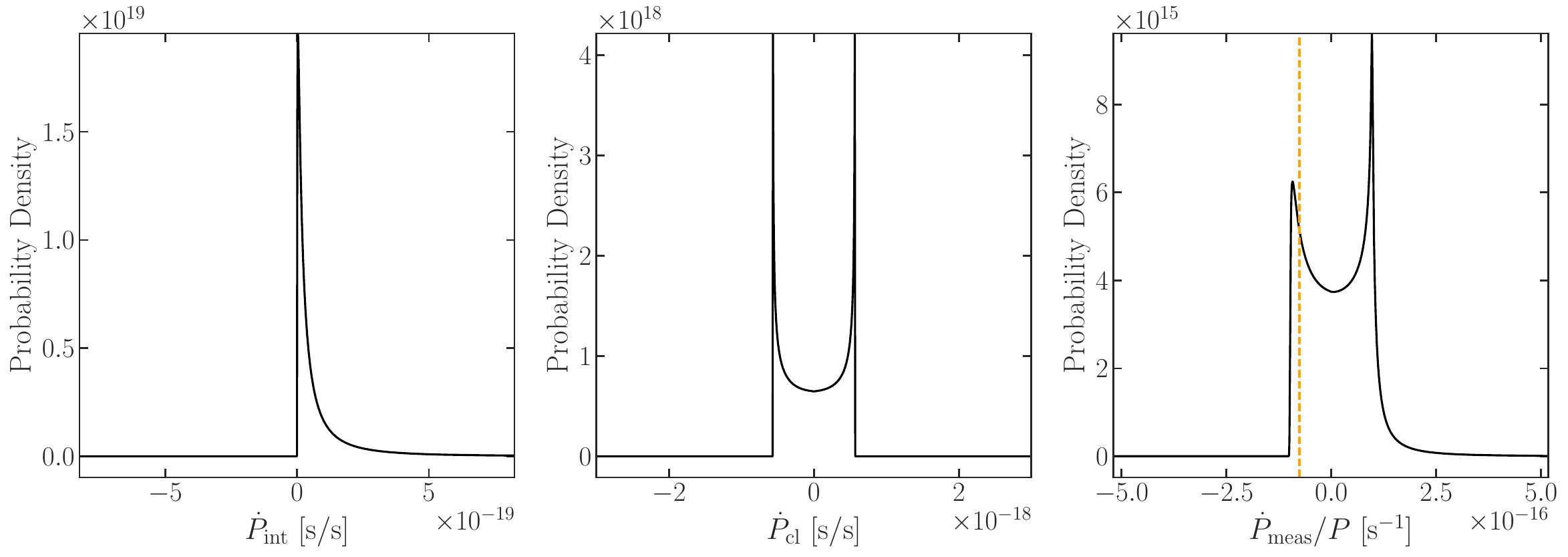}
    \caption{{\it Left panel:} The probability distribution of $\dot{P}_{\rm int}$ (intrinsic spin-down)
        for pulsar \emph{aa} in Terzan 5. Note that this distribution is of the same form as the one shown in the bottom panel of Figure \ref{fig:field_pulsars} but here shown on a linear scale. {\it Middle panel:} The probability distribution of $\dot{P}_{\rm cl}$ (due to the cluster's potential at the projected radius of pulsar \emph{aa} in a model fit to Terzan 5. {\it Right panel}: The convolution of the intrinsic and cluster $\dot{P}$
        distributions, then convolved with  a Gaussian distribution representing the uncertainty on the measured period derivative, transformed to \PdotP, with the observed value for pulsar \emph{aa} shown as a vertical dashed line.
    \label{fig:convolution}}
\end{figure*}

To combine these various effects into a likelihood function for $(\dot{P}/P)_{\rm meas}$ given a
model, we start with the distribution of $\dot{P}_{\rm cl}/P$ from the cluster's gravitational potential
and convolve with it the intrinsic distribution of $(\dot{P}/P)_{\rm int}$ values for the period of
a given pulsar. We additionally convolve the distribution with a Gaussian distribution, centered at
zero with a width equal to the uncertainty of $(\dot{P}/P)_{\rm meas}$ in order to fully incorporate
the uncertainty of the period derivative measurement. We then shift this distribution by the point
estimates for the contributions of the Galactic potential and the Shklovskii effect. This results in
a probability distribution for a measured period derivative which fully incorporates the physical
effects within the cluster, which depend on our model parameters as well as the effects of the
Galactic gravitational potential and the effects of the pulsars' proper motions. We use this
probability distribution to compute the likelihood of each measured period derivative. We show an
example of the combination of these various distributions into the final likelihood in Figure
\ref{fig:convolution}.

Many pulsars in GCs are in binary systems, and for systems with well-determined timing
solutions, the orbital period derivatives of these systems can be measured. The orbital period solutions are useful
because the orbital periods of these systems are of the order of days, while the intrinsic orbital decay of these systems acts over millions of years \citep[e.g.][]{Binney2008}. 
This means that, in the cases where the orbital period derivative can be measured, the changes in the orbital period can be entirely attributed to the acceleration from the cluster and the well-constrained effects of the Galactic potential and pulsar's proper motion. We note that some black widow pulsars, like pulsars \emph{J} and \emph{O} in 47\,Tuc, show orbital variability that is likely due to interactions with their companion \citep[e.g.][]{Shaifullah2016,Freire2017} which could be incorrectly interpreted as changes in the period derivative due to the cluster potential. The orbital period derivatives of these systems cannot be measured, so the these problematic systems are already excluded from our sample.

Due to the longer timescales and the
difficulties associated with determining the orbital periods of these systems (see
\citealt{Ridolfi2016} for details), the relative uncertainties on the orbital period derivatives are much larger than those of the spin-period derivatives. These
larger uncertainties mean that the likelihood functions for the observed orbital period
derivatives are wider and provide weaker constraints on the mass distribution of
the cluster, but we nonetheless use the orbital period derivatives of these systems
(pulsars \emph{E, H, I Q, R, S, T, U, X} and \emph{Y} in 47~Tuc
and pulsars \emph{ao, ap, au, av, aw} and \emph{ax} in Terzan~5\footnote{There are many additional pulsars in binary systems in Terzan 5
\citep[e.g.][]{Ransom2005} however the timing solutions for these systems lack reported uncertainties which are required for our method.
The spin-period timing solutions from \citet{Ransom2005} are similarly lacking reported uncertainties however, in practice, uncertainties
on spin-period solutions are so small that our method is insensitive to their value. In these cases we adopt a single value for the uncertainty
on the spin period derivatives, taking \SI{5e-21}{s/s} as a conservative estimate for these pulsars though we stress that our results are not 
sensitive to the adopted value.})
as an additional constraint on the cluster potential,
independent of any intrinsic effects on the period derivatives. We construct these likelihood functions in an identical way to the spin period
likelihoods but we neglect any effects intrinsic to the binary systems.

One avenue for future improvement of this methodology lies in including the dispersion measures of the
pulsars in the analysis. The dispersion measure of a pulsar provides a measure of the amount of
ionized gas between the pulsar and an observer, with a higher column density of free electrons producing a larger dispersion
measure. Given an estimate of the average dispersion measure between an observer and a cluster and a
model for the internal gas distribution within a cluster, the dispersion measures provide an
estimate of the line-of-sight position of each pulsar within the cluster.

For 47 Tuc, \citet{Abbate2018} used the pulsars within the cluster to infer the
internal gas distribution. These authors found that the pulsar data preferred a uniform gas distribution
within the cluster rather than a distribution that follows the stellar density (see also \citealt{Pancino2024a}), finding a gas density $n_{\rm g}$ of $0.23 \pm 0.05 \ \mathrm{cm^{-3}}$. This measurement, combined with the average cluster
dispersion measure $\mathrm{DM_{cl}}$ of $24.38 \pm 0.02 \ \mathrm{pc \ cm^{-3}}$ allows us to infer
the 3-D position of each pulsar within 47 Tuc, independent of our modeling. We describe the necessary modifications to equation \ref{eq:paz} in Appendix \ref{sec:DM-appendix}.

We implemented this alternative formulation and applied it to 47 Tuc to test if the dispersion measures
would enable the pulsar timing data to place stronger constraints on the mass distribution within our
models. We found that while the dispersion measures did in some cases provide stronger constraints from
individual pulsars, the uncertainties on the dispersion-measure-based line-of-sight positions are such that the overall constraints
are ultimately very similar to those provided by the density-based calculation described in equation~(\ref{eq:paz}).
Because the required internal gas models do not yet exist for Terzan~5 and because the dispersion measures
provided little to no improvement for 47 Tuc, we opt to simply use the density-based calculation for
both clusters for the remainder of this paper.

\subsection{Stellar mass functions}

The stellar mass function likelihoods are also Gaussian likelihoods, however, care must be taken
when extracting model values due to the non-trivial footprint of the observed \emph{HST} fields from which the mass functions were extracted. To ensure that we are extracting mass functions from the same corresponding regions in the models, we employ a Monte Carlo integration method which allows us to
handle the irregular overlapping \emph{HST} fields. This process is described in detail by
\citet{Dickson2023} and is implemented in the \gcfit package.

The only uncertainty formally included with the stellar mass function data is the Poisson counting
error. We introduce a nuisance parameter $F$ which scales up the uncertainties on the absolute
counts by a constant factor, leading to larger relative errors in regions with lower counts. This
error encapsulates additional sources of error that may not have been accounted for such as the
error associated with the conversion from luminosity to mass with an isochrone and the fact that
the mass function is being approximated as a broken power law, a functional
form which may not be a perfectly accurate representation of the true mass function of the cluster.

\subsection{Stellar populations and mass bins}
\label{sec:stellar-pops}

To generate the input mass bins for the \limepy models the \texttt{evolve{\_}mf} algorithm requires as inputs the mass function power-law slopes, as well as the age and metallicity of the stellar population. For 47 Tuc we adopt an age of 11.75 Gyr \citep{VandenBerg2013} and a metallicity of $\FeH = -0.72$ \citep[][2010 edition]{Harris1996}. Depending on the fit, as detailed below, we either allow the mass function power-law slopes to vary or fix the mass function.

Terzan 5 has been found to have at least three distinct stellar populations \citep{Ferraro2009,Origlia2013,
    Ferraro2016}. One of these populations, the most metal-poor population ($\FeH = -0.8$), only makes up a small
fraction of the cluster and can be neglected in our analysis. The other two population are a
young (4.5 Gyr) super-solar ($\FeH = 0.2$) population, making up about $40\%$ of the cluster (by
mass) and an old (12 Gyr) population with $\FeH = -0.2$ making up the other $60\%$ of
the cluster. With the young population making up a significant fraction of the cluster, we
cannot simply assume a single, old stellar population for Terzan 5 as we do for 47 Tuc as we want our remnant
populations to be as realistic as possible.

Tests with our stellar evolution algorithm show that this $40/60$ mixture of the young
and old stellar populations results in a remnant mass fraction around $35\%$ at the present day, made up of mostly white dwarfs. We can achieve a similar
remnant fraction ($\sim 34 \%$) with a single metal-poor ($\FeH = -0.2$), intermediate-age population of about 8 Gyr. Using this intermediate-age stellar population also
makes the main-sequence turnoff mass consistent with the maximum mass of main-sequence stars of $\sim 0.9 \ \Msun$ in our stellar mass function data, avoiding possible issues when comparing the observed mass function and model predictions.

While populations of different metallicities are expected to produce different remnants from similarly massive progenitors, this is a minor effect for our modeling. Our primary goal is to produce a realistic mix of remnants that together make up the correct fraction of the total mass of the cluster, a goal for which an intermediate-age population is a useful simplification.

Finally, the stellar mass function data available for Terzan 5 does not cover a wide enough range of
masses or distances from the cluster center to leave the mass function power-law slopes free while fitting as they are derived from a single field
and only extend down to $\sim 0.6 \ \Msun$. As such, we chose to fix the mass function slopes for Terzan
5 while still including the mass function data as a constraint on the visible stellar mass in three radial
bins between 0.67 and 1.67 arcminutes from the center.

We chose to adopt for the present-day mass function of \mbox{Terzan 5} the bottom-light IMF of \citet{Baumgardt2023b}, which was measured from star
clusters in the Milky Way and Magellanic Clouds and represents the best estimate for the IMF of massive
star clusters. We show in Section \ref{sec:results} that this mass function provides a satisfactory match to
the available stellar mass function data. This mass function has slopes of $\alpha_1 = 0.3, \alpha_2 = 1.65, \alpha_3 = 2.3$ and we again place our breakpoints at $0.5$ and $1$ \Msun.

\section{Results}
\label{sec:results}

To test the performance of our method, we run several fits for each cluster with different subsets of
the data introduced in Section \ref{sec:data}.

For 47 Tuc, we perform
three fits: (1) a fit to all available data for this cluster (\TucAllData), (2) a fit with the pulsar timing data held out (\TucNoPulsars), and (3) a fit to
the number density profile, pulsar timing data and a single field of mass function data\footnote{The field from $5.0-8.33$ arcmin, \HST proposal ID 11677. This field was chosen to roughly probe a similar radial region as the single field available for Terzan 5.} (\TucNoKin), designed
to emulate the data available for \mbox{Terzan 5}.

For Terzan 5 we also run three fits: (1) a fit with all of
the available data for this cluster (\TerAllData), (2) a fit with the pulsar data held out (\TerNoPulsars), and finally (3) a fit on just the number density
profile and pulsar data, with the kinematic data and mass function data held out (\TerNoKin), to test the reliability of the limited stellar kinematic data available for
\mbox{Terzan 5}.

These fits are summarized in Table \ref{tab:fitting-runs}. We present the median of the posterior probability distribution and $1\sigma$ credibility
intervals for the parameters of each of these fits in Table \ref{tab:fitting-results}\footnote{We have made the plots and sampler outputs for all six
of our fits available in an online repository: \href{https://zenodo.org/doi/10.5281/zenodo.12004419}{10.5281/zenodo.12004419}}.

\begingroup
\begin{table*}
    \centering
    \caption{Summary of the different model fits for 47 Tuc and Terzan 5, showing which datasets are included or held out in each case. The columns indicate if the models are fit to the number density profile (NDP), line-of-sight velocity dispersion profile (LOS), proper motion dispersion profile (PM), stellar mass function data (MF) and the pulsar timing data.
    \label{tab:fitting-runs}}

    \begin{tabular}{c | c c c c c}
        \hline
        {\centering Fit} & NDP    & LOS    & PM     & MF                 & Pulsars \\
        \hline
        \TucAllData      & \cmark & \cmark & \cmark & \cmark             & \cmark  \\
        \TucNoPulsars    & \cmark & \cmark & \cmark & \cmark             &         \\
        \TucNoKin        & \cmark &        &        & \cmark (one field) & \cmark  \\
        \hline
        \TerAllData      & \cmark & \cmark & \cmark & \cmark             & \cmark  \\
        \TerNoPulsars    & \cmark & \cmark & \cmark & \cmark             &         \\
        \TerNoKin        & \cmark &        &        &                    & \cmark  \\
        \hline
    \end{tabular}
\end{table*}
\endgroup

\begin{table*}
    \centering
    \caption{Medians and $1\sigma$ uncertainties of each model parameter for each of our fits.
        Entries without uncertainties indicate parameters that have been held fixed during fitting.
        We note that as discussed in \citet{Dickson2023}, the statistical uncertainties listed here
        likely underestimate the true uncertainties on each parameter and, in particular, our uncertainties
        on the cluster mass are likely closer to $10 \%$ \citep{Dickson2024a}. 
    \label{tab:fitting-results}}

    \begin{tabular}{c|c c c | c c c}
        \hline
        Cluster                                       & \textbf{\TucAllData}      & \TucNoPulsars & \TucNoKin & \textbf{\TerAllData} & \TerNoPulsars & \TerNoKin \\
        \hline

         $\hat{\phi}_0$ & \(6.08\substack{+0.08 \\ -0.08}\) & \(6.05\substack{+0.07 \\ -0.06}\) & \(6.00\substack{+0.10 \\ -0.06}\) & \(5.9\substack{+0.3 \\ -0.3}\) & \(6.0\substack{+1.3 \\ -0.3}\) & \(5.9\substack{+0.4 \\ -0.3}\) \\
         $M\ \left[10^6\ M_\odot\right]$ & \(0.899\substack{+0.006 \\ -0.006}\) & \(0.907\substack{+0.006 \\ -0.005}\) & \(0.96\substack{+0.02 \\ -0.01}\) & \(0.67\substack{+0.06 \\ -0.04}\) & \(0.79\substack{+0.06 \\ -0.07}\) & \(0.70\substack{+0.06 \\ -0.07}\) \\
         $r_{\mathrm{h}}\ \left[\mathrm{pc}\right]$ & \(6.68\substack{+0.04 \\ -0.04}\) & \(6.70\substack{+0.04 \\ -0.04}\) & \(7.03\substack{+0.06 \\ -0.06}\) & \(2.1\substack{+0.3 \\ -0.2}\) & \(2.0\substack{+0.4 \\ -0.3}\) & \(2.3\substack{+0.4 \\ -0.3}\) \\
         $\log_{10}\left(\hat{r}_{\mathrm{a}}\right)$ & \(1.73\substack{+0.05 \\ -0.03}\) & \(1.78\substack{+0.05 \\ -0.04}\) & \(4.88\substack{+2.01 \\ -1.88}\) & \(4.75\substack{+2.07 \\ -2.27}\) & \(5.39\substack{+1.77 \\ -2.50}\) & \(4.49\substack{+2.10 \\ -2.18}\) \\
         $g$ & \(1.50\substack{+0.03 \\ -0.03}\) & \(1.54\substack{+0.02 \\ -0.02}\) & \(1.50\substack{+0.02 \\ -0.03}\) & \(1.3\substack{+0.5 \\ -0.6}\) & \(2.0\substack{+0.2 \\ -0.5}\) & \(1.6\substack{+0.4 \\ -0.6}\) \\
         $\delta$ & \(0.47\substack{+0.01 \\ -0.01}\) & \(0.48\substack{+0.01 \\ -0.01}\) & \(0.489\substack{+0.008 \\ -0.013}\) & \(0.38\substack{+0.06 \\ -0.05}\) & \(0.34\substack{+0.06 \\ -0.03}\) & \(0.41\substack{+0.06 \\ -0.06}\) \\
         $s^{2}\ \left[\mathrm{arcmin^{-4}}\right]$ & \(0.0006\substack{+0.0003 \\ -0.0002}\) & \(0.0011\substack{+0.0028 \\ -0.0006}\) & \(0.0005\substack{+0.0002 \\ -0.0002}\) & \(7.84\substack{+4.87 \\ -5.18}\) & \(6.37\substack{+5.47 \\ -4.33}\) & \(7.33\substack{+5.13 \\ -4.80}\) \\
         $F$ & \(2.6\substack{+0.1 \\ -0.1}\) & \(2.59\substack{+0.09 \\ -0.10}\) & \(5.8\substack{+0.8 \\ -0.7}\) & \(1.9\substack{+0.3 \\ -0.2}\) & \(1.7\substack{+0.3 \\ -0.2}\) & \(1.9\substack{+0.3 \\ -0.2}\) \\
         $\alpha_{1}$ & \(0.38\substack{+0.03 \\ -0.02}\) & \(0.38\substack{+0.02 \\ -0.02}\) & 0.3 & 0.3 & 0.3 & 0.3 \\
         $\alpha_{2}$ & \(1.31\substack{+0.04 \\ -0.04}\) & \(1.31\substack{+0.04 \\ -0.04}\) & 1.65 & 1.65 & 1.65 & 1.65 \\
         $\alpha_{3}$ & \(2.23\substack{+0.03 \\ -0.03}\) & \(2.24\substack{+0.02 \\ -0.02}\) & 2.3 & 2.3 & 2.3 & 2.3 \\
         $\mathrm{BH}_{\mathrm{ret}}\ \left[\%\right]$ & \(0.28\substack{+0.05 \\ -0.04}\) & \(0.31\substack{+0.06 \\ -0.05}\) & \(0.31\substack{+0.21 \\ -0.10}\) & \(1.77\substack{+2.07 \\ -1.24}\) & \(5.35\substack{+2.90 \\ -2.96}\) & \(1.84\substack{+1.60 \\ -1.19}\) \\
         $d\ \left[\mathrm{kpc}\right]$ & \(4.41\substack{+0.02 \\ -0.01}\) & \(4.43\substack{+0.01 \\ -0.01}\) & \(4.45\substack{+0.03 \\ -0.03}\) & \(6.7\substack{+0.1 \\ -0.1}\) & \(6.7\substack{+0.1 \\ -0.1}\) & \(6.7\substack{+0.1 \\ -0.1}\) \\
        
        \hline
    \end{tabular}
\end{table*}

\subsection{47 Tuc}
\label{sec:47Tuc_results}

The \TucNoPulsars fit is very similar to the fit presented in \citet{Dickson2024a}, and we use this
fit as a baseline to evaluate the additional leverage provided by the pulsar data. The \TucAllData
fit to all of the available data is shown in Figure \ref{fig:47tuc-alldata-obs} and Figure
\ref{fig:47tuc-MF}, along with examples of the likelihood functions for the measured pulsar
period derivatives for the best-fitting models shown in Figure \ref{fig:Pdot_dists} (top panels) and all pulsars
in Figures \ref{fig:47Tuc-Paz-spin-grid1}, \ref{fig:47Tuc-Paz-spin-grid2} and \ref{fig:47Tuc-Paz-orbital-grid}.

\begin{figure*}
    \centering
    \includegraphics[width=0.8\linewidth]{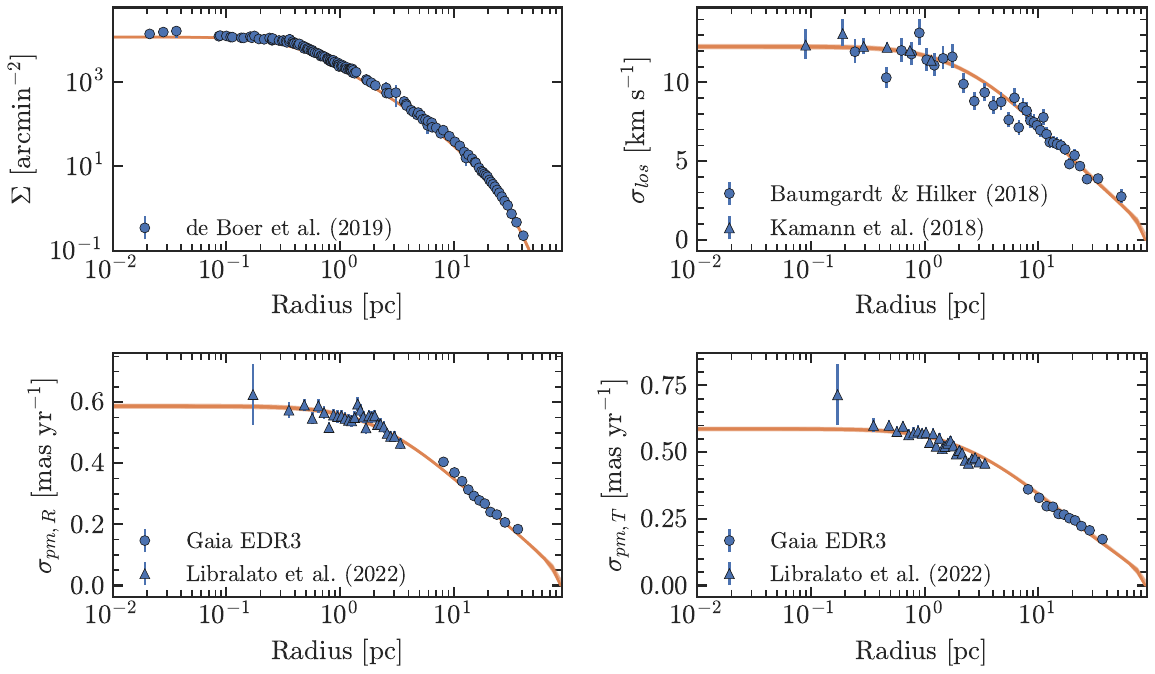}
    \caption{Best-fitting model (in orange) for the \TucAllData fit compared to different datasets
        (in blue). \emph{Top left:} Model fit to the projected number density profile. \emph{Top
            right:} Model fit to the projected line-of-sight velocity dispersion profile. \emph{Bottom
            panels:} Model fit to the projected proper motion dispersion profile, separated into radial
        (left panel) and tangential (right panel) components. The shaded regions represent the
        $1\sigma$ and $2\sigma$ credible intervals. 
    \label{fig:47tuc-alldata-obs}
    }
\end{figure*}

\begin{figure*}
    \centering
    \includegraphics[width=0.8\linewidth]{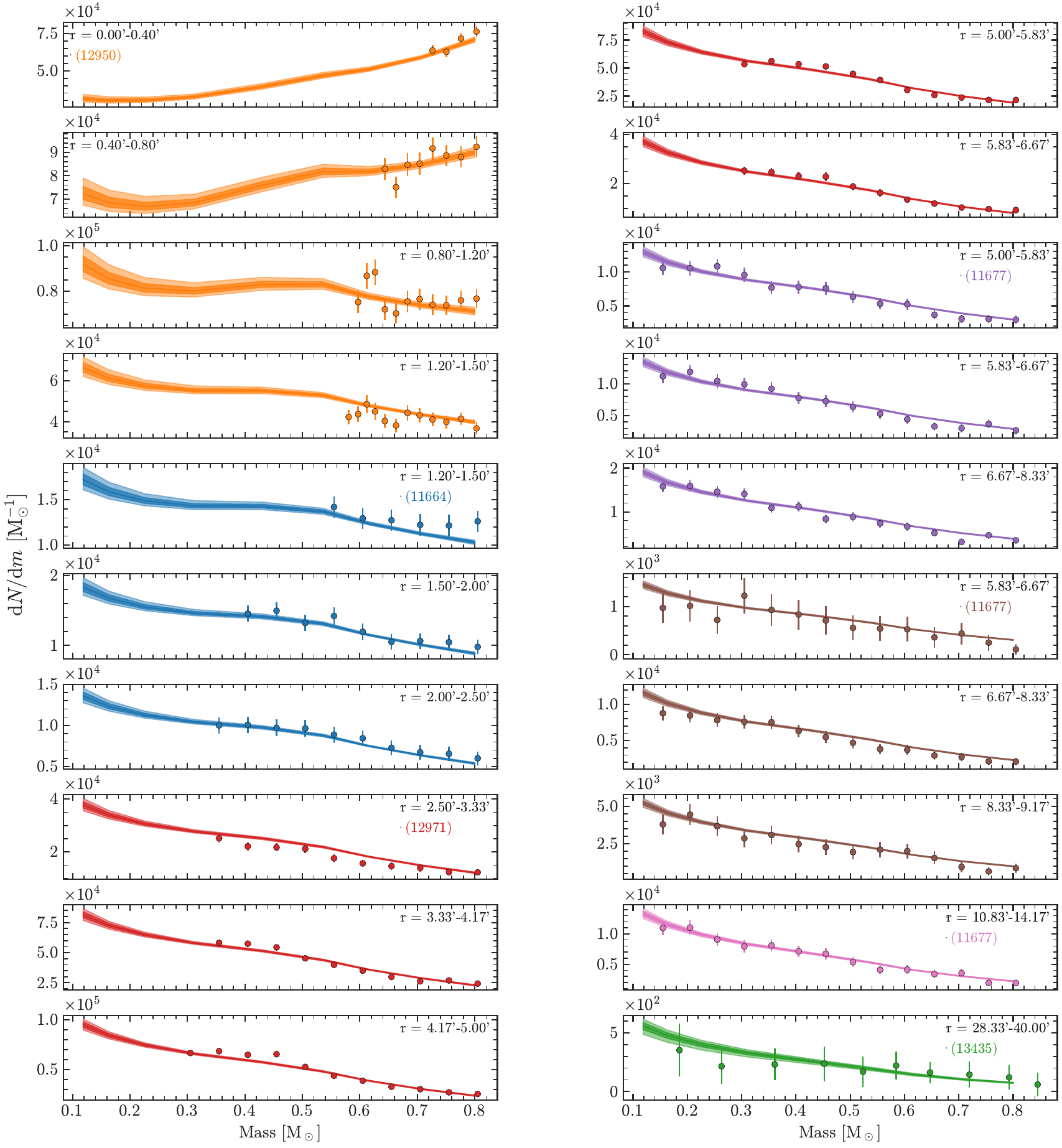}
    \caption{Continuation of Figure \ref{fig:47tuc-alldata-obs} showing the best-fitting model for
        the \TucAllData fit compared to the stellar mass function data of 47 Tuc. Each colour
        corresponds to a different \HST field and each panel to a different radial region within a
        field. We display the \HST proposal ID for in the first panel of each field.
    \label{fig:47tuc-MF}}
\end{figure*}

\begin{figure*}
    \centering
    \includegraphics[width=0.7\linewidth]{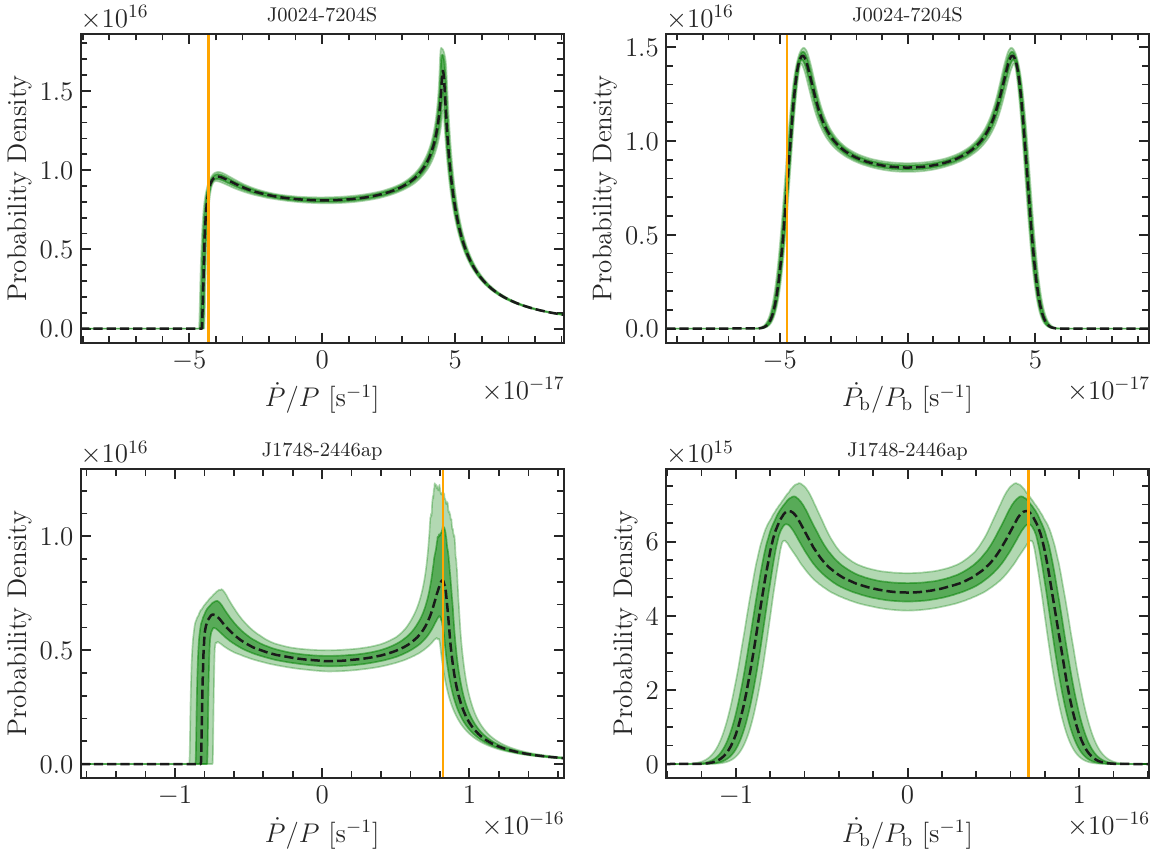}
    \caption{Likelihood functions corresponding to the best-fitting model for the observed \PdotP and \PbdotPb (recall that $P$ and $P_{\rm b}$ are the spin and orbital periods of the pulsars) for one pulsar in each cluster. {\it Top left:} The \PdotP likelihood for pulsar \emph{S} in 47 Tuc from the \TucAllData fit. {\it Top right:} The \PbdotPb likelihood for pulsar \emph{S} in 47 Tuc from the \TucAllData fit.
            {\it Bottom left:} The \PdotP likelihood  for pulsar \emph{ap} in Terzan 5 from the \TerAllData fit. {\it Bottom right:} The \PbdotPb likelihood for pulsar \emph{ap} in Terzan 5 from the \TerAllData fit. In each panel, we show the observed period derivative as a vertical orange line. We show similar plots for all of the pulsars in Figures \ref{fig:47Tuc-Paz-spin-grid1} through \ref{fig:Ter5-Paz-orbital-grid}.
    \label{fig:Pdot_dists}}
\end{figure*}

As an additional check on our fits, we compare the velocity dispersion of the pulsars to the prediction from our models.
For 47 Tuc, we limit this comparison to pulsars within $1 '$ from the center, which corresponds to the  isothermal (for pulsars) portion of our model. 
This leaves us with 22 pulsars for which we calculate a total proper motion dispersion of $0.37 \pm 0.10 \ \mathrm{mas} \ \mathrm{yr}^{-1}$.
The prediction from our \TucAllData model is $ 0.430 \pm 0.004 \ \mathrm{mas} \ \mathrm{yr}^{-1}$, in good agreement with the measured value.

An initial comparison of the \TucAllData and \TucNoPulsars fits reveals no significant differences,
either in model parameters, fit quality or derived quantities like the black hole mass fraction. We
take this agreement as an indication that the \TucNoPulsars fit already provides a very good
description of the underlying mass distribution and dynamics of the cluster, as probed by and fully
consistent with the pulsar data.

\begin{figure*}
    \centering
    \includegraphics[width=0.8\linewidth]{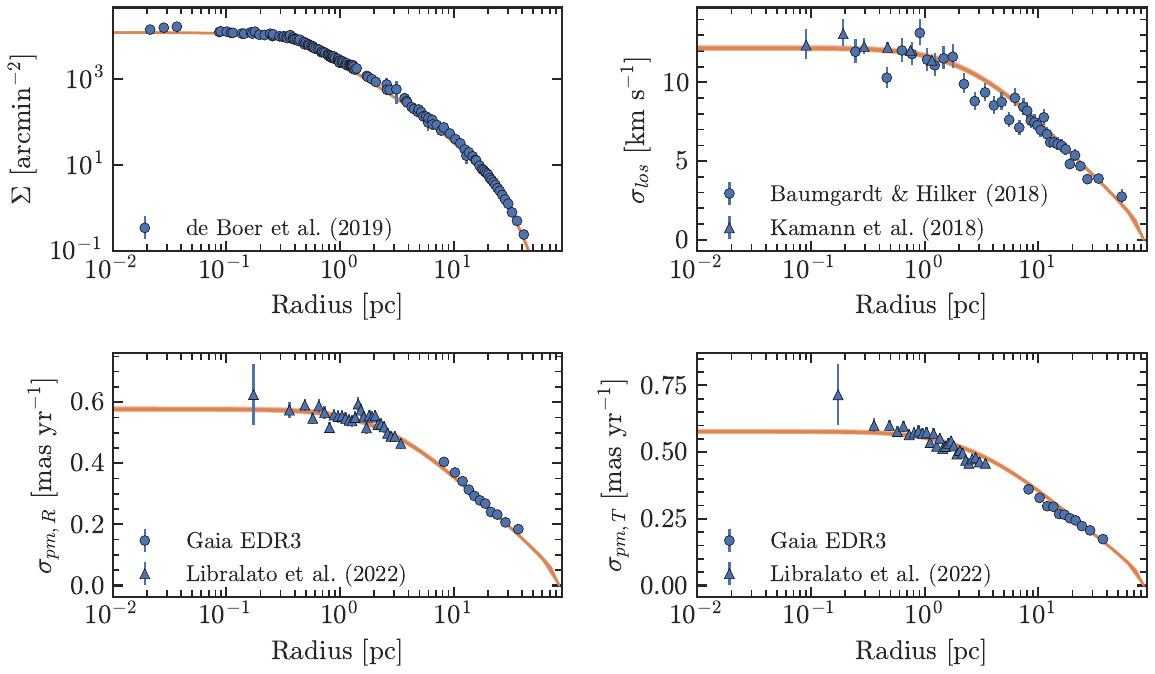}
    \caption{Same as Figure \ref{fig:47tuc-alldata-obs} but for the \TucNoKin fit. The best-fitting model is plotted along with the stellar kinematic data even though this data is excluded from the fit.
    \label{fig:47tuc-ter5sim-obs}}
\end{figure*}

Given the agreement between the \TucAllData and \TucNoPulsars fits, we turn to the third case in
order to evaluate the leverage provided by the pulsar data. In the \TucNoKin fit we seek to emulate
the data that we have for Terzan 5. For this fit, we fix the mass function to the bottom-light IMF
of \citet{Baumgardt2023b} discussed in Section \ref{sec:stellar-pops}. This mass function is a reasonable approximation for 47 Tuc and is similar to the best-fitting mass function we infer when the mass function is allowed to vary (see Table \ref{tab:fitting-results}). We show part of the
results for this third fit in Figure \ref{fig:47tuc-ter5sim-obs}, where the best-fitting model is
plotted along with the stellar kinematic data even though this data is excluded from the fit. This
model is in excellent agreement with the data, similar to the models that are directly fit on the
stellar kinematics and the best fit parameters for this fit are similar to the previous two fits (see Table \ref{tab:fitting-results}).
A comparison of the enclosed mass profiles of the \TucAllData and \TucNoKin fits reveal that the mass profiles vary by less than $\sim 5\%$ within the innermost \SI{1}{pc} (where the \TucNoKin fit contains less mass) and  the total mass varies only by $\sim 5\%$ with the \TucNoKin fit favoring a slightly higher mass.

\begin{figure}
    \centering
    \includegraphics[width=0.8\linewidth]{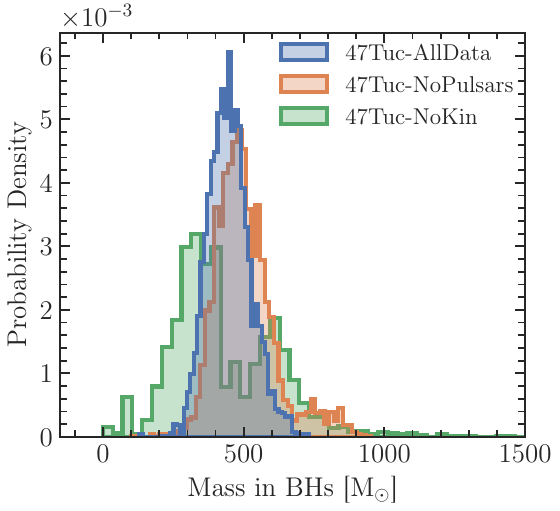}
    \caption{Posterior probability distribution of the mass in BHs in 47 Tuc for each of the fits
        summarized in Table \ref{tab:fitting-runs}.
    \label{fig:BHmass-47tuc}}
\end{figure}

We show our inferred posterior probability distribution for the cluster mass in BHs for each of our
three fits of 47 Tuc in Figure \ref{fig:BHmass-47tuc}. While the results for the \TucAllData and
\TucNoPulsars fits are quite similar and both resemble the results of \citet{Dickson2024a}, the
\TucNoKin fit is worth discussing in more detail. The most obvious feature of this fit is that the
posterior distribution of the mass in BHs is much broader than the cases with abundant stellar
kinematic and mass function data, and this posterior is also not uni-modal. Despite this, we can still place a very
stringent upper limit (99th~percentile) on the mass in BHs, limiting this mass to less than $\sim 0.1\%$
of the total cluster mass, even in the \TucNoKin fit. 
We further note that even though we have adopted a fixed IMF
for the \TucNoKin fit, the inferred BH content is consistent with the fits where we allow the IMF to vary.

These tests demonstrate that the pulsar timing data can provide
constraints on the central dynamics, BH content, and mass distribution of the cluster
very similar to the stellar kinematics, and in cases like Terzan 5 where the
stellar kinematic data are lacking, pulsar timing data may provide
an excellent substitute. With this in mind, we discuss the case of Terzan 5 next.

\subsection{Terzan 5}
\label{sec:Ter5_results}

We show the \TerAllData fit in Figures \ref{fig:Ter5-alldata-obs} and \ref{fig:Ter5-MF}, along with examples of the likelihood functions for the measured
period derivatives for the best-fitting models shown in Figure \ref{fig:Pdot_dists} (bottom panels) and all the pulsars in Figures \ref{fig:Ter5-Paz-spin-grid1}, \ref{fig:Ter5-Paz-spin-grid2} and \ref{fig:Ter5-Paz-orbital-grid}. Our
best-fitting model for the \TerAllData fit is in excellent agreement with the data and is
able to fully reproduce all of the observables, including the pulsar data. The comparative lack of
data for Terzan 5 means that our inferred model parameters for Terzan 5 generally
have larger uncertainties compared to our fits of 47 Tuc. We present the median and $1\sigma$ uncertainties on all model parameters in Table \ref{tab:fitting-results}.

\begin{figure*}
    \centering
    \includegraphics[width=0.8\linewidth]{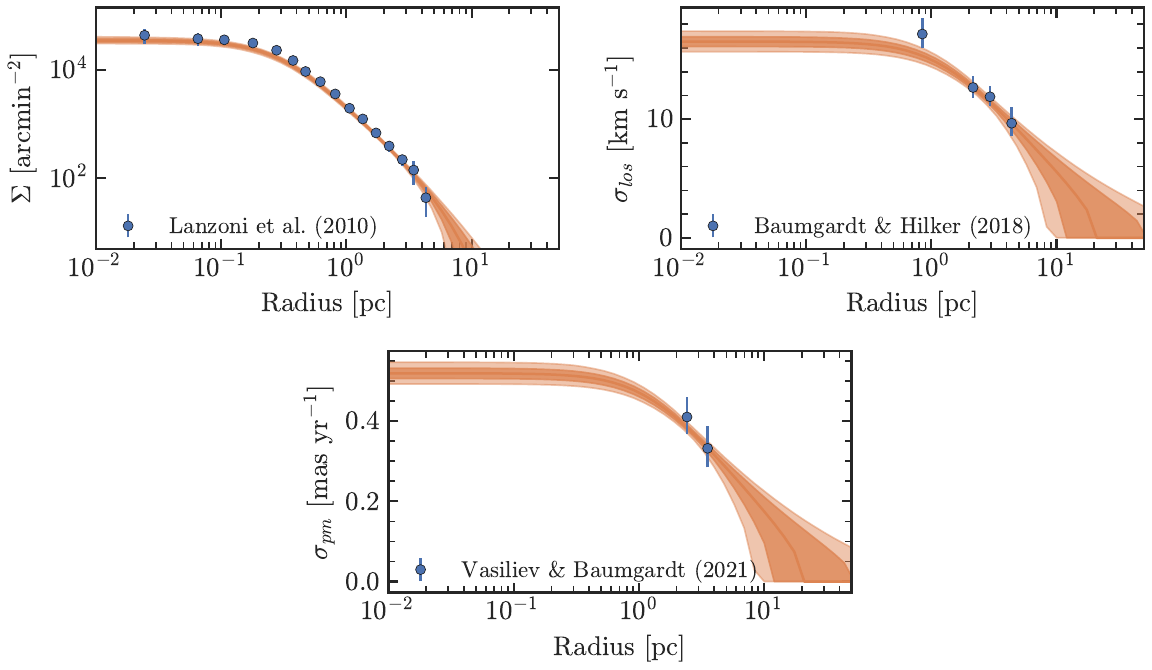}
    \caption{Best-fitting model (in orange) for the \TerAllData fit compared to different datasets
        (in blue). \emph{Top left:} Model fit to the projected number density profile. \emph{Top
            right:} Model fit to the projected line-of-sight velocity dispersion profile. \emph{Bottom:}
        Model fit to the projected proper motion dispersion profile. The shaded regions represent
        the $1\sigma$ and $2\sigma$ credible intervals of the model fits.
    \label{fig:Ter5-alldata-obs}}
\end{figure*}

For Terzan 5, most pulsars do not have reported proper motions, and fewer still have both components
of the proper motions reported, meaning that we have insufficient data to calculate a velocity dispersion
from the pulsar data to compare to our model prediction.

When we do not fit on the pulsar data (\TerNoPulsars),
the uncertainties on model parameters and related quantities generally become
larger than was the case for the \TerAllData fit. This comparison suggests that
the pulsar data can play a
more dominant role in constraining the models in the case of Terzan 5 (compared to 47 Tuc), given
the lack of stellar kinematic data for this bulge cluster.

The third fit, \TerNoKin, was done to test the reliability of the stellar
kinematics data given the challenge of membership determination in the Galactic bulge. As we find
for 47 Tuc, even when we exclude the stellar kinematic data from the fit, the pulsar timing data
provides enough constraints that the resulting best-fitting model is relatively similar
and generally in good agreement with the held-out data. With the insight from 47 Tuc (see Section
\ref{sec:47Tuc_results}) that pulsars can provide strong constraints on the mass distribution of a
cluster, we interpret this agreement as a sign that the existing stellar kinematics for
Terzan 5, while sparse, are likely not suffering from significant contamination
or other systematic effects.

As mentioned previously, the stellar mass function data available for Terzan 5 does not cover a wide
enough range of masses or radii to a sufficient completeness level to allow us to leave the mass function slopes as free parameters when
fitting models. We adopted the IMF of \citet{Baumgardt2023b} for each of our three
fits and we show in Figure \ref{fig:Ter5-MF} that this mass function is in excellent agreement with
the available data.

\begin{figure}
    \centering
    \includegraphics[width=0.8\linewidth]{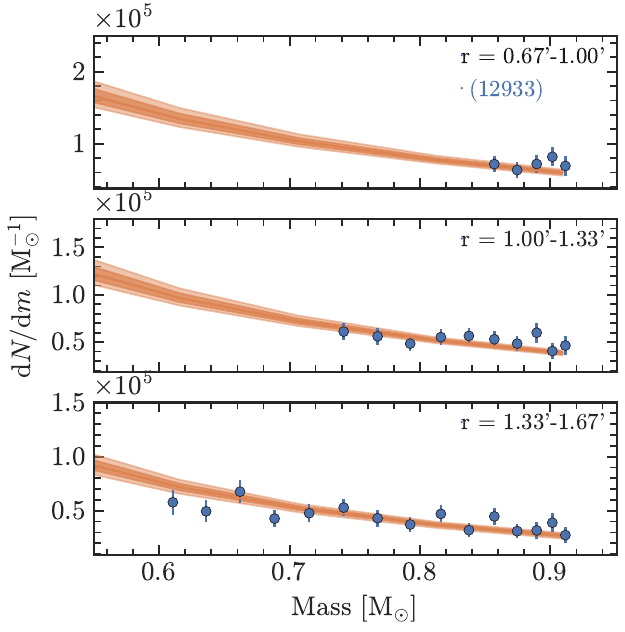}
    \caption{Continuation of Figure \ref{fig:Ter5-alldata-obs} showing the best-fitting model for
        the \TerAllData fit compared to the stellar mass function data of Terzan 5. Each panel
        corresponds to a different radial region within the \HST field.
    \label{fig:Ter5-MF}}
\end{figure}

\begin{figure}
    \centering
    \includegraphics[width=0.8\linewidth]{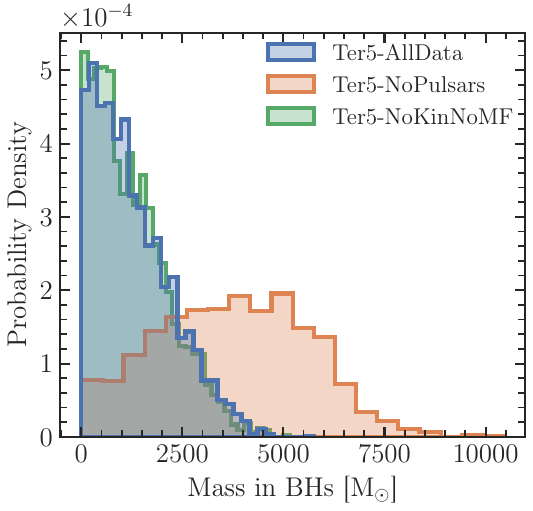}
    \caption{Posterior probability distribution of the mass in BHs in Terzan 5 for each of the fits
        summarized in Table \ref{tab:fitting-runs}.
    \label{fig:BHmass-ter5}}
\end{figure}

We show our inferred cluster masses in BHs for each of our fits of Terzan 5 in Figure
\ref{fig:BHmass-ter5}. Comparing the fits, it is again obvious that the pulsar data is providing
most of the constraining power, with the \TerNoPulsars fit resulting in an
upper bound on the mass in BHs roughly double that of the two fits that do include the pulsar data.
Even with the inclusion of the pulsar data we are only able to place an upper limit on the mass in
BHs in this cluster, though we do significantly improve on existing estimates, lowering the range of
allowed masses by a factor of $\sim 10$ (see discussion in \mbox{Section \ref{sec:BH-comp}}). 
With these results, we cannot rule out the possibility that Terzan~5 contains zero BHs at the present day and indeed our posterior distribution of
mass in BHs is peaked towards zero for both fits that include the pulsar data.

\section{Discussion}
\label{sec:discussion}

\subsection{Constraints from pulsars}

\begin{figure}
    \centering
    \includegraphics[width=0.8\linewidth]{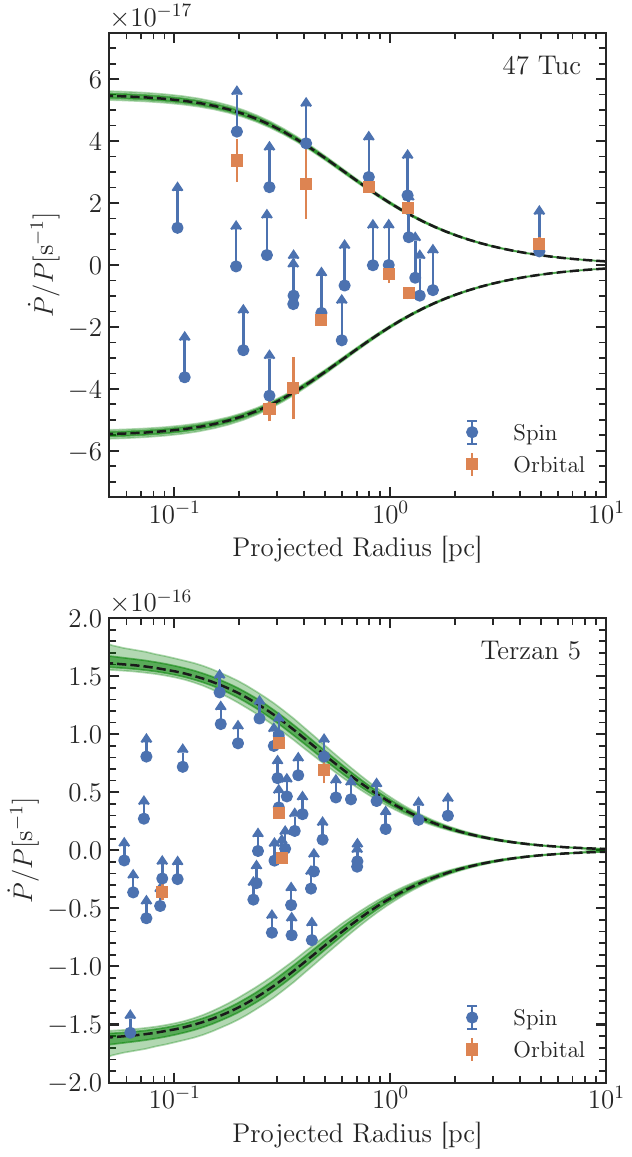}
    \caption{Minimum and maximum values of \PdotP allowed from the
        acceleration in the cluster potential for the best-fitting models of our \TucAllData (top) and \TerAllData
        (bottom) fits. We show the median values as a dashed line while the shaded regions represent
        the $1\sigma$ and $2\sigma$ credible intervals of the model fits. The observed \PdotP of the pulsars are shown with blue circles and the observed \PbdotPb
        are shown with orange squares. Uncertainties on the \PbdotPb points represent measurement errors while
        the upward-facing error bars on the \PdotP points show the typical width of the $\dot{P}_{\rm int}$ distribution
        at each pulsar's location in the $P$-$\dot{P}$ plane shown in Figure \ref{fig:field_pulsars}. Note that pulsars
        above the maximum contour are not disallowed by our models due to the long tails on the spin-down side of the
        likelihood function shown in Figure \ref{fig:convolution}. We note that we have converted the projected radius
        of each pulsar from angular to linear units using the median value of our inferred distances to each cluster
        from the \texttt{AllData} fits (see Table \ref{tab:fitting-results}).
        \label{fig:trumpets}}
\end{figure}

The specific constraints provided by the pulsar data are on the local acceleration of each pulsar,
and thus on the enclosed mass profile of the cluster. The most stringent constraints on the cluster 
gravitational field come from pulsars with large \emph{negative} observed values of \PdotP.
This can be seen most easily in the rightmost panel of Figure \ref{fig:convolution} where the
positive side of the \PdotP distribution has a long tail due to the convolution with the intrinsic
spin-down distribution. By contrast, the negative side of the \PdotP distribution truncates sharply to zero at the \PdotP corresponding to the maximum possible acceleration at a given projected radius. In terms of specific pulsars, this means that pulsar \emph{S} and \emph{aa}
in 47 Tuc provide the strictest limits on the enclosed mass while pulsar \emph{ae} provides the
strictest limits in Terzan 5. When inferring the dark remnant content of a cluster (in particular BHs), the
ideal pulsars would be very close to the center of the cluster and would have large negative
observed values of \PdotP\footnote{These pulsars would fall on the far side of the cluster along the line of sight. Pulsars on the near side of the cluster would have positive line-of-sight accelerations due to the cluster potential which would shift the observed \PdotP to positive values where our method is less constraining because intrinsic spin-down also shifts the observed \PdotP towards positive values.}, as these pulsars place the strongest constraints on the central mass
distribution of their host cluster, where the mass density is dominated by heavy remnants due to mass segregation. We show in Figure \ref{fig:trumpets} the minimum and maximum
allowed values of \PdotP and \PbdotPb due to the cluster potential, where pulsars providing these
stronger constraints fall along the bottom contour of the allowed values of \PdotP.

Our finding of pulsar \emph{S} being the most constraining pulsar in 47~Tuc is not new and was discussed by \citet{Giersz2011}. These authors reported that when attempting to find a suitable
Monte Carlo model of 47~Tuc, the tension between the central surface brightness and the large
negative acceleration of pulsar \emph{S} was the single most impactful factor. Similarly, Figure 2
(extended) of \citet{Klzlltan2017} shows that pulsar \emph{S} is just barely compatible with their model
and could potentially be a driving factor in requiring more mass in the cluster center, therefore
favoring models with an IMBH.
Pulsars 47Tuc-\emph{S} and Ter5-\emph{ae} do have a binary companions \citep{Freire2017, Prager2017a}, however,
due to the fact that the companions are low-mass WDs any spin-up effects from accretion which would shift the
observed spin period derivatives towards more negative values are very unlikely. Pulsar 47Tuc-\emph{aa} appears
to be an isolated pulsar, leaving little possibility for accretion-induced
spin-up. Because these pulsars are not likely to experience any spin period change
from accretion, the constraints that their spin period derivatives place on the mass distribution of
the cluster are likely trustworthy.

While pulsars with large negative observed values of \PdotP provide the strongest constraints, the
majority of pulsars have values that are much closer to zero or even on the positive extreme of the
distribution. These pulsars nonetheless provide useful constraints on the mass distribution of the
cluster, particularly when it comes to constraining the total mass of the cluster. These pulsars
have observed values of \PdotP that would be technically compatible with any model containing some
minimum mass at their radius. As the mass of the model grows however, the range of possible \PdotP
values grows, lowering the probability of observing these less extreme values. In this way the
pulsars provide not just a minimum enclosed mass at their projected radius, but also some leverage on the
exact value of the enclosed mass. We can see this when we compare our fits of Terzan 5 with
different subsets of the data, as the best-fitting models found when including the pulsar data in the fits
(\TerAllData and \TerNoKin) are less massive than the fit
that excludes the pulsar data (\TerNoPulsars).

\subsection{Mass of Terzan 5}

\begin{figure}
    \centering
    \includegraphics[width=0.9\linewidth]{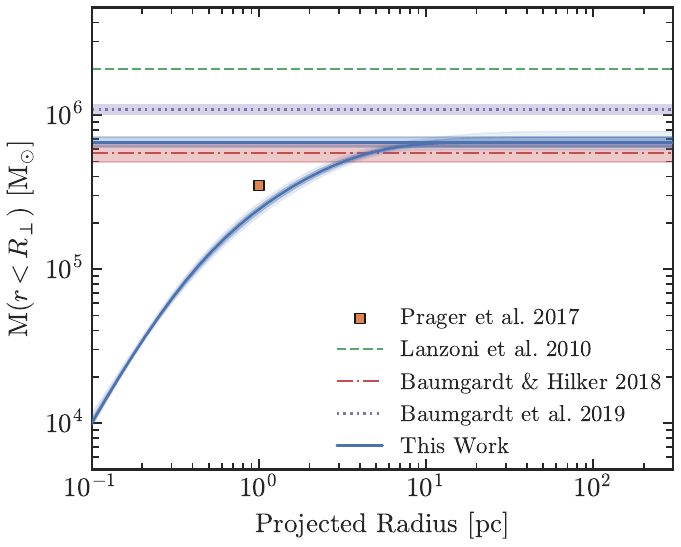}
    \caption{The inferred cumulative mass profile of Terzan~5 from our \TerAllData fit.
        We show the median values as a solid line (blue) while the shaded regions represent
        the $1\sigma$ and $2\sigma$ credible intervals of the model fit.
        The enclosed mass at \SI{1}{pc} measured by \citet{Prager2017a} is shown with an orange square. We show
        our inferred total mass and the values from \citet{Lanzoni2010}, \citet{Baumgardt2018}
        and \citet{Baumgardt2019b} as horizontal dashed lines where the shaded regions represent
        the $1\sigma$ credible intervals.
    \label{fig:ter5-mass-comp}}
\end{figure}

The total mass of Terzan 5 is somewhat uncertain in the current literature, with mass estimates based on
photometry \citep[e.g.][]{Lanzoni2010} a factor of a few higher than those based on kinematics and dynamical modeling
\citep[e.g.][]{Baumgardt2018, Prager2017a}. We show in Figure \ref{fig:ter5-mass-comp} our inferred
cumulative mass profile along with several literature values for the cluster mass. Our profile is in good agreement with
the estimate of the enclosed mass at \SI{1}{pc} from \citet{Prager2017a} and our total mass estimate
of $0.67\substack{+0.06 \\ -0.04} \times 10^6 \ \Msun$ is in good agreement with the total masses of
\citet{Baumgardt2018} and \citet{Baumgardt2019b} given that the true uncertainty on our inferred mass from multimass modeling is
likely closer to $10 \%$ \citep[see Section 3 in][]{Dickson2024a}. The masses inferred from kinematics and dynamical modeling, including our own
value, are a factor of $2-3$ times smaller than the mass inferred by \citet{Lanzoni2010} from
photometry. A lower present-day mass for Terzan 5 potentially has many important implications, in particular
for studies that seek to model the star-formation and chemical enrichment histories of
this system \citep[e.g.][]{Romano2023}.

\subsection{Comparison of mass in BHs to literature results}
\label{sec:BH-comp}

Our models allow us to place strong constraints on the mass in BHs in both 47 Tuc and
Terzan 5. We compare our results with other studies that investigate the BH population
in these clusters in Table \ref{tab:BH-content}.

47 Tuc is a well-studied cluster with many previous works investigating its BH content, using a
variety of methods. In general, we see that recent works which employ Monte Carlo models \citep{Weatherford2019, Ye2022a} generally
infer larger BH populations (with larger uncertainties) despite taking very different approaches while works employing equilibrium models
favor somewhat smaller populations of BHs. Our models place an upper limit (99th percentile) on the total mass in BHs in 47 Tuc
of $649 \ \Msun$, very similar the results of \citet{Dickson2024a} who did not consider pulsar timing data. Our results are also in very good agreement with the upper limit of $578 \ \Msun$ on the mass of a central IMBH reported by \citet{DellaCroce2024}.
These authors employ action-based distribution function models to derive an upper limit on the mass
of a putative IMBH in 47 Tuc, which as discussed earlier (see discussion in Section \ref{sec:intro}), is expected to have similar dynamical effects to
a compact central cluster of BHs. Our inferred upper limit on the mass in BHs corresponds to an
upper limit of $0.07\%$ on the BH mass fraction \fbh, slightly lower than the upper limit of
$0.09\%$ found by \citet{DellaCroce2024}\footnote{Our larger mass in BHs corresponds to a smaller \fbh because we infer a larger total mass for 47 Tuc with our multimass models than \citet{DellaCroce2024} with their single-mass models, in line with the tendency for single-mass equilibrium models to underestimate the total cluster mass \citep[see][]{Henault-Brunet2019}.}. These upper limits significantly limit the room for a more massive
potential IMBH in 47 Tuc, which we discuss in the following section.

Terzan 5 on the other hand, has only a single estimate of its BH
content reported in the literature, by \citet{Prager2017a}. These authors, also fitting on pulsar
data, test for the presence of an IMBH in the center of 47 Tuc, reporting marginal evidence of a
$\sim 500 \ \Msun$ IMBH and an upper limit of \SI{30000}{\Msun}. Our posterior for the mass in BHs in
Terzan 5 is peaked near zero and the 99th~percentile upper limit is $3860 \ \Msun (\fbh < 0.6\%)$,
representing an improvement of nearly one order of magnitude on the existing constraints on the mass in BHs.

\begingroup
\begin{table}
\setlength{\tabcolsep}{2pt}
    \centering
    \caption{Reported masses in BHs (or upper limits) in 47~Tuc and Terzan~5 from dynamical studies in the  literature and from this work.
        Note that we have scaled the BH mass fraction reported by \citet{Weatherford2019} to our inferred total mass to
        facilitate comparison. We note that the uncertainties on the black hole content reported by studies that employ
        \limepy models (\citealt{Henault-Brunet2020, Dickson2024a} and this work) are likely underestimated by a factor
        of $2-3$ (see discussion in \citealt{Dickson2024a}). Listed uncertainties correspond to the $1\sigma$ uncertainties.
\label{tab:BH-content}}
    \begin{tabular}{l l l}

        \hline
        Study                                              & 47 Tuc                & Terzan 5              \\
        \hline
        H\'enault-Brunet et al. (2020)\tablenotemark{\rm \dag} & $430 \substack{+386                           \\ -301} \ \Msun$     &  \\
        Weatherford et al. (2020)                          & $1037 \substack{+1640                         \\ -922} \ \Msun$     &   \\
        Ye et al. (2022)                                   & $\sim 2375 \ \Msun$   &                       \\
        Della Croce et al. (2024)\tablenotemark{\rm *}         & $<578 \ \Msun$        &                       \\
        Dickson et al. (2024)                              & $420\substack{+150                            \\ -80}\ \Msun$ & \\
        This work                                          & $446 \substack{+75                            \\ -72} \ \Msun$ & $<3860 \ \Msun$ \\
        Prager et al. (2017)\tablenotemark{\rm *}              &                       & $< 30 \, 000 \ \Msun$ \\
        \hline
    \end{tabular}

    \tablenotetext{\dag}{ Note that the the posterior of \citet{Henault-Brunet2020} is peaked towards zero.}
    \tablenotetext{*}{These are upper limits on the mass of a putative IMBH, not the total mass in BHs.}

\end{table}
\endgroup

\subsection{An IMBH in 47 Tuc?}

GCs have long been suggested to host IMBHs. For proposed detections in multiple clusters using
various techniques, see e.g. \citet{Gerssen2002}, \citet{Noyola2008}, \citet{Jalali2012},
\citet{Lutzgendorf2015}, \citet{Kamann2016}, \citet{Baumgardt2017a}, \citet{Klzlltan2017}, \citet{Paduano2024}, and \citet{Haberle2024b}. 
Many of these possible detections have however been contested. Typically, follow-up
studies find that the dynamical effects of a central IMBH and other ingredients like a large
population of centrally concentrated dark stellar remnants (including stellar-mass black holes -
BHs) are highly degenerate, making the detection of a non-accreting IMBH very difficult \citep[see
    e.g.][]{McNamara2003, VanDerMarel2010, Freire2017, Zocchi2017, Zocchi2019, Gieles2018, Baumgardt2019, Mann2019,Mann2020,
    Henault-Brunet2020}. Ruling out the presence of an IMBH is also very difficult, with most works
placing only an upper limit on the dark mass in the core of a cluster \citep[e.g.][]{Mann2019,Haberle2021,DellaCroce2024}.
\citet{Haberle2024b} find seven stars with velocities above the inferred escape velocity in the inner 0.08 pc of \omegacen,
and they conclude that this implies the presence of an IMBH with a mass $\gtrsim8200\,\Msun$. We note that only two of those seven stars
are above the escape velocity of the mass model of \citet{Dickson2023}, but they are the two closest to the cluster center and remain difficult to reconcile with explanations other than an IMBH,
making this the strongest case for an IMBH in a Galactic GC so far. 

Particularly relevant to this work are the results from \citet{Klzlltan2017, Kzltan2017a} and
\citet{Paduano2024} who each report evidence for an IMBH in 47~Tuc. \citet{Klzlltan2017} reported
evidence for a $2300 \ \Msun$ IMBH in the center of 47\,Tuc on the basis of pulsar timing measurements.
In order to constrain the mass of a central IMBH, they compared snapshots from a grid of \Nbody models
(with and without an IMBH) to surface density and velocity dispersion profiles. The best-fitting models
with and without an IMBH were then compared to measurements of pulsar accelerations (based on period derivatives)
due to the cluster's gravitational potential, and the set of models with a central IMBH was found to be more consistent
with these measurements.

Critically, several follow-up studies identified limitations and possible issues with the analysis
of \citet{Klzlltan2017} (see \citealt{Freire2017, Mann2019,Mann2020,Henault-Brunet2020}). Among the issues
raised was the assumption of a short cluster distance of $\SI{4}{kpc}$\footnote{The latest estimates
    of the distance to 47\,Tuc based on \Gaia data place it $4.52 \pm 0.03~ \mathrm{kpc}$ away
    \citep{Baumgardt2021}.}, the lack of primordial binaries  in their \Nbody models, and the use of a
grid of isolated \Nbody models that have a much steeper present-day mass function than the
bottom-light mass function that is observed for 47\,Tuc, all of which could affect the inferred
amount of dark mass in the cluster core.

We note that models of 47~Tuc that do not include a central IMBH have been shown to satisfyingly
reproduce its velocity dispersion profile, number density profile, and stellar mass function data
\citep{Baumgardt2019a, Henault-Brunet2020, Dickson2023, Dickson2024a}. These studies, however, either
did not consider the pulsar data, or they checked that the models were consistent with the pulsars'
maximum accelerations and projected radial distribution but did not directly incorporate the pulsar
timing data in the fitting process.

Our results allow us to address the claims of \citet{Klzlltan2017} since we are directly fitting our
models to the pulsar timing data. As mentioned
previously, our fit to all the available 47 Tuc data (\TucAllData) yields an upper limit of $649 \
    \Msun$ in BHs \mbox{($\sim 0.07 \%$ of the total cluster mass)}. This argues against the conclusions
of \citet{Klzlltan2017} that the pulsar data requires a central IMBH of $2300 \ \Msun$ to explain the
observed values of \PdotP. We performed a fit of 47 Tuc with the distance set to \SI{4}{kpc}, resulting in a
noticeably worse quality of fit but also a larger ($\sim 2400 \ \Msun$) population
of BHs, perhaps indicating that the low distance assumed by \citet{Klzlltan2017} contributes to the differences
in our inferred dark mass in 47 Tuc.

The potential IMBH detection reported by \citet{Paduano2024} has a much wider range of possible masses,
with the nominal 1$\sigma$ uncertainty range spanning \mbox{$56 - 6000 \ \Msun$}. 
The results of \citet{DellaCroce2024}, who infer an upper limit on the mass of an IMBH in 47 Tuc of $578 \ \Msun$
reduce this range substantially. The fact that our inferred upper limit on the mass in BHs is consistent with the 
upper limit on the mass of an IMBH from \citet{DellaCroce2024} highlights the fact there is very little room for a 
significant dark mass in the center of 47 Tuc.

While the effects of an IMBH and a centrally concentrated population of BHs are expected to be similar, they are not
identical because the dark mass is more centrally concentrated if it is in an IMBH. This means that an upper limit on
the mass in BHs does not necessarily correspond directly to the upper limit on the mass of an IMBH. Despite this, we
expect that in our models, these effects are likely degenerate due to the spatial regions probed by our datasets.
The half-mass radius of the BH population in our \TucAllData fit is $\sim 0.1 \ \mathrm{pc}$, which is more centrally
concentrated than the majority of the pulsars in this cluster. While we do have a few pulsars inside this radius, the
pulsars with large negative period derivatives (those that provide the most stringent constraints on the enclosed mass)
are located outside of this radius meaning we are largely insensitive to the spatial extent and concentration of the mass
in BHs in the central regions of the cluster.

We note that given the central line-of-sight velocity dispersion of 47 Tuc, the radius of influence of an IMBH of $\sim 600 \ \Msun$
is $\lesssim 0.02 \ \mathrm{pc}$ which at the distance of 47~Tuc corresponds to $\lesssim 0.8''$, a much smaller region than is probed
by either the pulsar data or the stellar kinematic data.
This means that any future work seeking to use stellar kinematics to understand the nature of the dark mass in the core of this cluster 
would need to precisely measure the velocities of individual stars in the central arcsecond of 47
Tuc.

\subsection{Central velocity dispersion of Terzan 5}

The existing stellar kinematic data for
Terzan 5 does not probe the central regions of the cluster, but the pulsar timing data allows us to
independently predict the central velocity distribution of bright stars, without directly relying on
any stellar kinematics. This also allows us to validate the existing observed stellar kinematics for a cluster
like Terzan 5 where bulge contamination makes membership selection especially difficult. Our
predicted central dispersion for Terzan 5 based on fit \TerNoKin is  $15.7 \substack{+0.5 \\ -0.4} \ \textrm{km s}^{-1}$.
Future work seeking to constrain the central kinematics of this cluster can use this prediction as a benchmark
to which new measurements can be compared. For example, a cusp in the central velocity dispersion within the radius
of influence of an IMBH should manifest as a velocity dispersion larger than this predicted central velocity dispersion
given that our models do not contain an IMBH.

\subsection{Terzan 5: a comparison with \omegacen}

The formation of Terzan 5 and its multiple populations is a topic of much debate in the literature,
with suggested formation channels ranging from the stripped core of an accreted dwarf galaxy
\citep{Ferraro2009, Ferraro2016} to a surviving fragment of the proto-bulge \citep{Ferraro2009,
    Ferraro2016, Taylor2022} to the product of a collision between a typical GC and a giant molecular cloud or young
massive cluster \citep{McKenzie2018, Pfeffer2021, Bastian2022}.

While it is generally accepted that the metal-rich population in Terzan 5 is too enriched to have
formed in a low-mass dwarf galaxy \citep[e.g.][]{Bastian2022}, we can further investigate the
dynamical evolution of this object by comparing Terzan 5 to \omegacen,
a well-studied GC which is frequently suggested to be the core of an accreted dwarf galaxy
\citep[e.g.][]{Pfeffer2021}.
\omegacen is the most massive GC in the Milky Way \citep[e.g.][2010 edition]{Harris1996} and
hosts multiple stellar populations with a large spread in iron abundance
\citep[e.g.][]{Johnson2010a, Bellini2017}. Several studies have presented dynamical models of
\omegacen which suggest that the cluster is host to a very large population of black holes
\citep[e.g.][]{Zocchi2019, Baumgardt2019, Dickson2024a} typically making up about $5\%$ of the total
cluster mass (consistent with having retained almost all its BHs). The fact that we infer a much
smaller population of BHs and indeed rule out a population
larger than $0.6 \%$ of the total cluster mass in Terzan~5 suggests a different evolutionary history for
this object. We note that \omegacen is more
massive than Terzan~5 by a factor of $\sim 3-5$, and has a factor of $\sim 5$ larger $r_{\rm h}$,
which both contribute to a higher retention of BHs by the present day for \omegacen through an order
of magnitude longer relaxation time (diminishing the importance of dynamical ejections of BHs throughout
the evolution of the cluster). Additionally, the higher metallicity of Terzan 5 compared to \omegacen
is expected to produce lower-mass BHs from the same initial progenitor masses
\citep[e.g.][]{Fryer2012,Banerjee2020}, increasing the number of BHs lost to natal kicks and reducing
the fraction of the cluster mass in BHs resulting from a given IMF.

While the origin of Terzan 5 remains uncertain, the discussion above highlights that further studies 
of the evolution of Terzan 5 along with its BH population could help to shed light on the initial 
conditions and formation of this peculiar system.

\section{Conclusions}
\label{sec:conclusion}

In this work, we presented a method to directly fit multimass dynamical models of GCs
to pulsar timing data, allowing to infer the mass distribution of clusters. We applied our method to
47 Tuc, a well-studied cluster with a wealth of conventional stellar kinematic and local stellar mass
function data as well as a large population of pulsars.
We use this cluster as a benchmark by which we evaluate the performance of our method.
We then applied our method to Terzan~5, a bulge cluster host to the largest population of pulsars of
any Milky Way GC and lacking in conventional stellar kinematic data. Our main conclusions are as
follows:

\begin{enumerate}
    \item For clusters like 47 Tuc and Terzan 5 that are host to large populations of pulsars, the
          timing solutions of these pulsars can place similar constraints on the mass distribution
          and dynamics of their host cluster as conventional stellar kinematics. We demonstrate that
          the pulsar timing data allows us to accurately predict held-out stellar kinematic data and place
          strong constraints on the BH content of clusters.

    \item We infer new and improved values for the mass and structural parameters of Terzan 5, finding a total
          mass of $0.67\substack{+0.06 \\ -0.04} \times 10^6 \ \Msun$, and a (3D) half-mass radius of
          \(2.1\substack{+0.3 \\ -0.2} \ \mathrm{pc} \). This mass is consistent with other dynamical
          estimates but is smaller by a factor of $2-3$ than the estimate derived from photometry.

    \item We refine existing constraints on the BH content of 47~Tuc and lower the existing upper limit
          on the mass in BHs of Terzan~5 by an order of magnitude. We infer the presence
          of $446 \substack{+75 \\ -72} \ \Msun$ in BHs in 47~Tuc and place an upper limit on the mass
          in BHs in Terzan~5 of  $3860 \ \Msun$.

    \item Our results do not support the $\sim2300 \ \Msun$ IMBH reported by
          \citet{Klzlltan2017} in the center of 47 Tuc on the basis of pulsar timing data, as we instead infer
          an upper limit of $649 \ \Msun$ in BHs in this cluster, representing $\sim 0.07 \%$ of the total cluster
          mass. This adds to several follow-up studies that refuted this original claim, but it is the first time
          that pulsar timing data, a crucial component of the \citet{Klzlltan2017} study, is revisited as a direct
          constraint on the dynamical models.

    \item We predict the central velocity dispersion of Terzan~5, independently of any stellar
          kinematic data, finding a dispersion of $15.7 \substack{+0.5 \\ -0.4} \ \mathrm{km \ s^{-1}}$.
          This prediction provides a baseline to which future work seeking to measure the central
          kinematics of this cluster can be compared.
\end{enumerate}

The next generation of radio telescopes is expected to dramatically increase the number of detected
pulsars in GCs \citep[e.g.][]{Hessels2015, Ridolfi2021, Chen2023a, Berteaud2024}, allowing us to
apply our methodology to clusters that are not currently known to host large numbers of
pulsars.

Perhaps the most promising candidate to host a large
number of undiscovered pulsars is Liller 1, a bulge cluster that is qualitatively very similar to
Terzan 5 \citep[e.g.][]{Ferraro2021}. \mbox{Liller 1} is a massive cluster ($\sim 1 \times 10^6 \
  \Msun$, \citealt{Baumgardt2019b}) that is similarly compact to Terzan~5
and has been found to have a similarly high stellar encounter rate \citep{Saracino2015a}. This high
stellar encounter rate, combined with strong gamma-ray emission detected from this cluster, suggests
that it may be host to hundreds of undiscovered MSPs \citep{Tam2011}.

Like Terzan 5, Liller 1 is difficult to observe due to its location, with bulge contamination and
strong differential reddening \citep[e.g.][]{Pallanca2021} making the collection of conventional
stellar kinematic data very difficult. As we have shown for Terzan 5, pulsars present a unique
opportunity to investigate the internal dynamics of even heavily obscured clusters.

\pagebreak
\section*{Acknowledgments}
% squeeze this into the left column
\vspace{-6 mm}

\begin{acknowledgments}

    We thank the anonymous referee for helpful comments and suggestions that improved the quality of this work.
    
    We thank Barbara Lanzoni for kindly sharing with us the number density profile of Terzan 5 presented
    in \citet{Lanzoni2010}.

    VHB acknowledges the support of the Natural Sciences and Engineering Research Council of Canada
    (NSERC) through grant RGPIN-2020-05990, and a New Faculty Grant from the Faculty of Graduate Studies
    and Research of Saint Mary's University.
    ND is grateful for the support of the Durland Scholarship in Graduate Research.
    MG acknowledges financial support from the grants PID2021-125485NB-C22, EUR2020-112157, CEX2019-000918-M funded 
    by MCIN/AEI/10.13039/501100011033 (State Agency for Research of the Spanish Ministry of Science and Innovation) and SGR-2021-01069 (AGAUR).

    This work made extensive use of Paulo Freire's database of pulsars in GCs
    (\url{https://www3.mpifr-bonn.mpg.de/staff/pfreire/GCpsr.html}).
    
    This research was enabled in part by support provided by ACENET (\url{www.ace-net.ca}) and the
    Digitial Research Alliance of Canada (\url{https://alliancecan.ca}).

\end{acknowledgments}


\vspace{5mm}
% \facilities{HST(STIS), Swift(XRT and UVOT), AAVSO, CTIO:1.3m,
%     CTIO:1.5m,CXO}


\software{   
    astropy \citep{TheAstropyCollaboration2013,TheAstropyCollaboration2018, TheAstropyCollaboration2022}, 
    dynesty \citep{Speagle2020, Koposov2023}.
    emcee \citep{Foreman-Mackey2013,Foreman-Mackey2019},
    gala \citep{Price-Whelan2017},
    matplotlib \citep{Hunter2007},
    numpy \citep{Harris2020},
    pandas \citep{Reback2020},
    scipy \citep{Virtanen2020},
    seaborn \citep{Waskom2021}
}



\appendix

\section{Supplementary Material}

\subsection{Incorporating dispersion measures}
\label{sec:DM-appendix}

Given a description of the gas distribution within a cluster \citep[e.g.][]{Abbate2018} Equation \ref{eq:paz} can be modified to use the dispersion measure of a pulsar to form the
line-of-sight probability distribution instead of using the model line-of-sight density profile:

\begin{equation}
	P\left(a_{{\rm cl},z} \mid R_i\right) \propto \frac{\mathrm{d} m}{\mathrm{d} a_{{\rm cl},z}}=\frac{\mathrm{d} m}{\mathrm{d} z} \abs{\frac{\mathrm{d} z}{\mathrm{d} a_{{\rm cl},z}}}=
	\frac{P\left( z \mid z_{{\rm DM},i}, \sigma_{z, {\rm DM},i}\right)}{\left|\frac{\mathrm{d} a_{{\rm cl},z}}{\mathrm{d} z}\right|},
	\label{eq:paz-DM}
\end{equation}
where we have replaced the $\frac{\mathrm{d} m}{\mathrm{d} z}$ term with $P\left( z \mid z_{\rm DM}, \sigma_{z, {\rm
			DM}}\right)$, the probability distribution of line-of-sight positions $z$ given the predicted
line-of-sight position from the dispersion measure $z_{\rm DM}$ and its uncertainty $\sigma_{z, {\rm
				DM}}$. The values of $z_{{\rm DM},i}$ and $\sigma_{z, {\rm DM}}$ can be calculated from the an
individual dispersion measure $\mathrm{DM}_i$, $n_g$ and $\mathrm{DM_c}$ as $z_{{\rm DM},i} = ({\rm
			DM}_i - {\rm DM_c})/n_g$ (under the assumption of a uniform gas distribution) with $\sigma_{z, {\rm DM}}$ following from Gaussian error propagation.
$P\left( z \mid z_{{\rm DM},i}, \sigma_{z, {\rm DM},i}\right)$ is then a Gaussian
probability density function, centered at $z_{{\rm DM},i}$ with a dispersion of
$\sigma_{z, {\rm DM}}$:

\begin{equation}
	P\left( z \mid z_{{\rm DM},i}, \sigma_{z, {\rm DM},i}\right) = \frac{1}{\sigma_{z, {\rm DM},i}\sqrt{2\pi}}
	\exp\left( -\frac{1}{2}\left(\frac{z-z_{{\rm DM},i}}{\sigma_{z, {\rm DM},i}}\right)^{\!2}\,\right).
\end{equation}

\subsection{Pulsar Data}

\begin{table*}

	\centering
	\caption{Pulsar timing data used in this work for 47 Tuc. The columns indicate the pulsar ID, projected radius, spin period, spin-period derivative, the uncertainty on the spin-period derivative, the orbital period, the orbital period derivative, the uncertainty on the orbital period derivative, and the proper motion in R.A. and Dec. with associated uncertainties. In the rightmost column we indicate the work in which the timing solution was derived.
		Timing References: F17: \citet{Freire2017}, R16: \citet{Ridolfi2016}, FR18: \citet{Freire2018}.
	\label{tab:pulsars_47tuc}}

	\begin{tabular}{lrrrrrrrrrc}
		\hline
		Pulsar ID    & $R$ & $P$  & $\dot{P}$ & $\Delta \dot{P}$  & $ P_{\rm b}$  & $\dot{P}_{\rm b}$  & $\Delta \dot{P}_{\rm b}$  & $\mu_{\alpha^*}$& $\mu_{\delta}$ & Ref. \\

         & [arcmin] & [ms] & [s/s] & [s/s] & [days] & [s/s] & [s/s] & $[\mathrm{mas} \ \mathrm{yr}^{-1}]$ & $[\mathrm{mas} \ \mathrm{yr}^{-1}]$ & \\
		\hline

        J0024-7204C  & 1.2298  & 5.75678  & -4.99e-20       & 2e-24    & --                  & --         & --      & $5.2   \pm  0.1   $ & $-3.1   \pm  0.1 $  & F17  \\
        J0024-7204D  & 0.6483  & 5.35757  & -3.42e-21       & 9e-25    & --                  & --         & --      & $4.24  \pm  0.07  $ & $-2.24  \pm  0.05 $ & F17  \\
        J0024-7204E  & 0.6205  & 3.53633  & 9.85e-20        & 5e-25    & 2.2568483           & 4.8e-12    & 2e-13   & $6.15  \pm  0.03  $ & $-2.35  \pm  0.06 $ & F17  \\
        J0024-7204F  & 0.2149  & 2.62358  & 6.45e-20        & 7e-25    & --                  & --         & --      & $4.52  \pm  0.08  $ & $-2.50  \pm  0.05 $ & F17  \\
        J0024-7204G  & 0.2781  & 4.04038  & -4.22e-20       & 2e-24    & --                  & --         & --      & $4.5   \pm  0.1   $ & $-2.9   \pm  0.1 $  & F17  \\
        J0024-7204H  & 0.7677  & 3.21034  & -1.83e-21       & 1e-24    & 2.357696895         & -7e-13     & 6e-13   & $5.1   \pm  0.2   $ & $-2.8   \pm  0.2 $  & F17  \\
        J0024-7204I  & 0.2772  & 3.48499  & -4.59e-20       & 2e-24    & 0.2297922489        & -8e-13     & 2e-13   & $5.0   \pm  0.2   $ & $-2.1   \pm  0.2 $  & F17  \\
        J0024-7204J  & 1.0185  & 2.10063  & -9.79e-21       & 9e-25    & --                  & --         & --      & $5.27  \pm  0.06  $ & $-3.59  \pm  0.09 $ & F17  \\
        J0024-7204L  & 0.1627  & 4.34617  & -1.22e-19       & 1e-24    & --                  & --         & --      & $4.4   \pm  0.2   $ & $-2.4   \pm  0.2 $  & F17  \\
        J0024-7204M  & 1.0688  & 3.67664  & -3.84e-20       & 5e-24    & --                  & --         & --      & $5.0   \pm  0.3   $ & $-2.0   \pm  0.4 $  & F17  \\
        J0024-7204N  & 0.4793  & 3.05395  & -2.19e-20       & 2e-24    & --                  & --         & --      & $6.3   \pm  0.2   $ & $-2.8   \pm  0.2 $  & F17  \\
        J0024-7204O  & 0.0806  & 2.64334  & 3.03e-20        & 6e-25    & --                  & --         & --      & $5.01  \pm  0.05  $ & $-2.58  \pm  0.08 $ & F17  \\
        J0024-7204Q  & 0.9502  & 4.03318  & 3.4e-20         & 6e-25    & 1.1890840496        & -1e-12     & 2e-13   & $5.2   \pm  0.1   $ & $-2.6   \pm  0.1 $  & F17  \\
        J0024-7204R  & 0.1519  & 3.48046  & 1.48e-19        & 3e-24    & 0.06623147751       & 1.9e-13    & 4e-14   & $4.8   \pm  0.1   $ & $-3.3   \pm  0.2 $  & F17  \\
        J0024-7204S  & 0.215   & 2.83041  & -1.21e-19       & 1e-24    & 1.2017242354        & -4.9e-12   & 4e-13   & $4.5   \pm  0.1   $ & $-2.5   \pm  0.1 $  & F17  \\
        J0024-7204T  & 0.3179  & 7.58848  & 2.94e-19        & 1e-23    & 1.126176771         & 2.5e-12    & 1.1e-12 & $5.1   \pm  0.6   $ & $-2.6   \pm  0.7 $  & F17  \\
        J0024-7204U  & 0.9386  & 4.34283  & 9.52e-20        & 2e-24    & 0.42910568324       & 6.6e-13    & 5e-14   & $4.6   \pm  0.2   $ & $-3.8   \pm  0.1 $  & F17  \\
        J0024-7204W  & 0.087   & 2.35234  & -8.66e-20       & 1e-24    & --                  & --         & --      & $6.1   \pm  0.5   $ & $-2.6   \pm  0.3 $  & R16  \\
        J0024-7204X  & 3.828   & 4.77152  & 1.84e-20        & 7e-25    & 10.921183545        & 6e-12      & 2e-12   & $5.8   \pm  0.1   $ & $-3.3   \pm  0.2 $  & R16  \\
        J0024-7204Y  & 0.3743  & 2.19666  & -3.52e-20       & 8e-25    & 0.5219386107        & -8.2e-13   & 7e-14   & $4.4   \pm  0.1   $ & $-3.4   \pm  0.1 $  & F17  \\
        J0024-7204Z  & 0.1506  & 4.55445  & -4.56e-21       & 1e-22    & --                  & --         & --      & $4     \pm  2     $ & $1      \pm  2 $    & F17  \\
        J0024-7204aa & 0.465   & 1.84538  & -4.59e-20       & 1.5e-23  & --                  & --         & --      & $4.6   \pm  0.8   $ & $-4.6   \pm  1.3 $  & FR18 \\
        J0024-7204ab & 0.2092  & 3.70464  & 9.82e-21        & 8e-24    & --                  & --         & --      & $4.2   \pm  0.6   $ & $-2.9   \pm  0.5 $  & F17  \\

		\hline
	\end{tabular}

\end{table*}

\begin{table*}

	\centering
	\caption{Same as Table \ref{tab:pulsars_47tuc} but for Terzan 5.
		Timing References: L00: \citet{Lyne2000}, R05: \citet{Ransom2005}, P17: \citet{Prager2017a}, C18: \citet{Cadelano2018}, A18: \citet{Andersen2018}, R21: \citet{Ridolfi2021}, P24: \citet{Padmanabh2024a}.
	\label{tab:pulsars_ter5}
	}

	\begin{tabular}{lrrrrrrrc}
		\hline
		Pulsar ID    & $R$ & $P$  & $\dot{P}$ & $\Delta \dot{P}$  & $ P_{\rm b}$  & $\dot{P}_{\rm b}$  & $\Delta \dot{P}_{\rm b}$ & Ref. \\

         & [arcmin] & [ms] & [s/s] & [s/s] & [days] & [s/s] & [s/s] & \\
		\hline

            J1748-2446C  & 0.179 & 8.4361   & -6.06e-19    & 4e-21 & --      & --       & --      & L00              \\
            J1748-2446D  & 0.693 & 4.71398  & 1.3e-19      & --    & --      & --       & --      & R05              \\
            J1748-2446E  & 0.361 & 2.1978   & -1.8e-20     & --    & --      & --       & --      & R05              \\
            J1748-2446F  & 0.125 & 5.54014  & 4e-21        & --    & --      & --       & --      & R05              \\
            J1748-2446G  & 0.185 & 21.6719  & 3.9e-19      & --    & --      & --       & --      & R05              \\
            J1748-2446H  & 0.227 & 4.92589  & -8.3e-20     & --    & --      & --       & --      & R05              \\
            J1748-2446I  & 0.03  & 9.57019  & -7.1e-20     & --    & --      & --       & --      & R05              \\
            J1748-2446J  & 0.948 & 80.3379  & 2.5e-18      & --    & --      & --       & --      & R05              \\
            J1748-2446K  & 0.22  & 2.96965  & -9.4e-20     & --    & --      & --       & --      & R05              \\
            J1748-2446L  & 0.149 & 2.2447   & -1.7e-20     & --    & --      & --       & --      & R05              \\
            J1748-2446M  & 0.083 & 3.56957  & 4.9e-19      & --    & --      & --       & --      & R05              \\
            J1748-2446N  & 0.154 & 8.6669   & 5.5e-19      & --    & --      & --       & --      & R05              \\
            J1748-2446O  & 0.119 & 1.67663  & -6.9e-20     & --    & --      & --       & --      & R05              \\
            J1748-2446Q  & 0.36  & 2.812    & -3.6e-20     & --    & --      & --       & --      & R05              \\
            J1748-2446R  & 0.101 & 5.02854  & 4.7e-19      & --    & --      & --       & --      & R05              \\
            J1748-2446S  & 0.249 & 6.11664  & 6.4e-20      & --    & --      & --       & --      & R05              \\
            J1748-2446T  & 0.443 & 7.08491  & 3.1e-19      & --    & --      & --       & --      & R05              \\
            J1748-2446U  & 0.148 & 3.28914  & 3e-19        & --    & --      & --       & --      & R05              \\
            J1748-2446V  & 0.178 & 2.07251  & -9.5e-20     & --    & --      & --       & --      & R05              \\
            J1748-2446W  & 0.037 & 4.20518  & 1.2e-19      & --    & --      & --       & --      & R05              \\
            J1748-2446X  & 0.488 & 2.99926  & 5.9e-20      & --    & --      & --       & --      & R05              \\
            J1748-2446Y  & 0.056 & 2.04816  & 1.5e-19      & --    & --      & --       & --      & R05              \\
            J1748-2446Z  & 0.033 & 2.46259  & -8.6e-20     & --    & --      & --       & --      & P17              \\
            J1748-2446aa & 0.222 & 5.78804  & -4.4e-19     & --    & --      & --       & --      & P17              \\
            J1748-2446ab & 0.038 & 5.11971  & 4.2e-19      & --    & --      & --       & --      & P17              \\
            J1748-2446ac & 0.337 & 5.08691  & 2.3e-19      & --    & --      & --       & --      & P17              \\
            J1748-2446ae & 0.032 & 3.65859  & -5.7e-19     & --    & --      & --       & --      & P17              \\
            J1748-2446af & 0.145 & 3.30434  & -2.3e-19     & --    & --      & --       & --      & P17              \\
            J1748-2446ag & 0.167 & 4.44803  & 1.2e-20      & --    & --      & --       & --      & P17              \\
            J1748-2446ah & 0.127 & 4.96515  & 5.7e-19      & --    & --      & --       & --      & P17              \\
            J1748-2446ai & 0.192 & 21.22838 & 1.4e-18      & --    & --      & --       & --      & P17              \\
            J1748-2446aj & 0.17  & 2.95891  & 1.41232e-19  & 6e-24 & --      & --       & --      & C18              \\
            J1748-2446ak & 0.287 & 1.8901   & 8.8495e-20   & 6-24  & --      & --       & --      & C18              \\
            J1748-2446am & 0.044 & 2.93382  & -1.368e-19   & 3e-23 & --      & --       & --      & A18              \\
            J1748-2446an & 0.201 & 4.802    & 1.55746e-19  & 6-24  & --      & --       & --      & R21              \\
            J1748-2446ao & 0.156 & 2.27438  & 8.6979e-20   & 1e-24 & 57.5556 & 1.65e-10 & 9e-12   & P24              \\
            J1748-2446ap & 0.253 & 3.74469  & 3.07e-19     & 1e-24 & 21.3882 & 1.3e-10  & 2.1e-11 & P24              \\
            J1748-2446aq & 0.038 & 12.52194 & -7.16198e-19 & 6e-24 & --      & --       & --      & P24              \\
            J1748-2446as & 0.084 & 2.32646  & 2.559829e-19 & 6e-25 & --      & --       & --      & P24              \\
            J1748-2446at & 0.123 & 2.18819  & -5.89966e-20 & 4e-25 & --      & --       & --      & P24              \\
            J1748-2446au & 0.053 & 4.54822  & -1.06797e-19 & 2e-24 & --      & --       & --      & P24              \\
            J1748-2446av & 0.045 & 1.84945  & -4.25047e-20 & 2e-25 & 3.38166 & -1e-11   & 2e-12   & P24              \\
            J1748-2446aw & 0.156 & 13.04908 & 1.306465e-18 & 3e-24 & 0.73138 & 5.92e-12 & 2e-14   & P24              \\
            J1748-2446ax & 0.161 & 1.9435   & -9.5495e-21  & 7e-25 & 30.2088 & -1.3e-11 & 9e-12   & P24              \\

		\hline
	\end{tabular}

\end{table*}

\subsection{Supplementary Figures}

\begin{figure*}
\centering
	\includegraphics[width=0.8\textwidth]{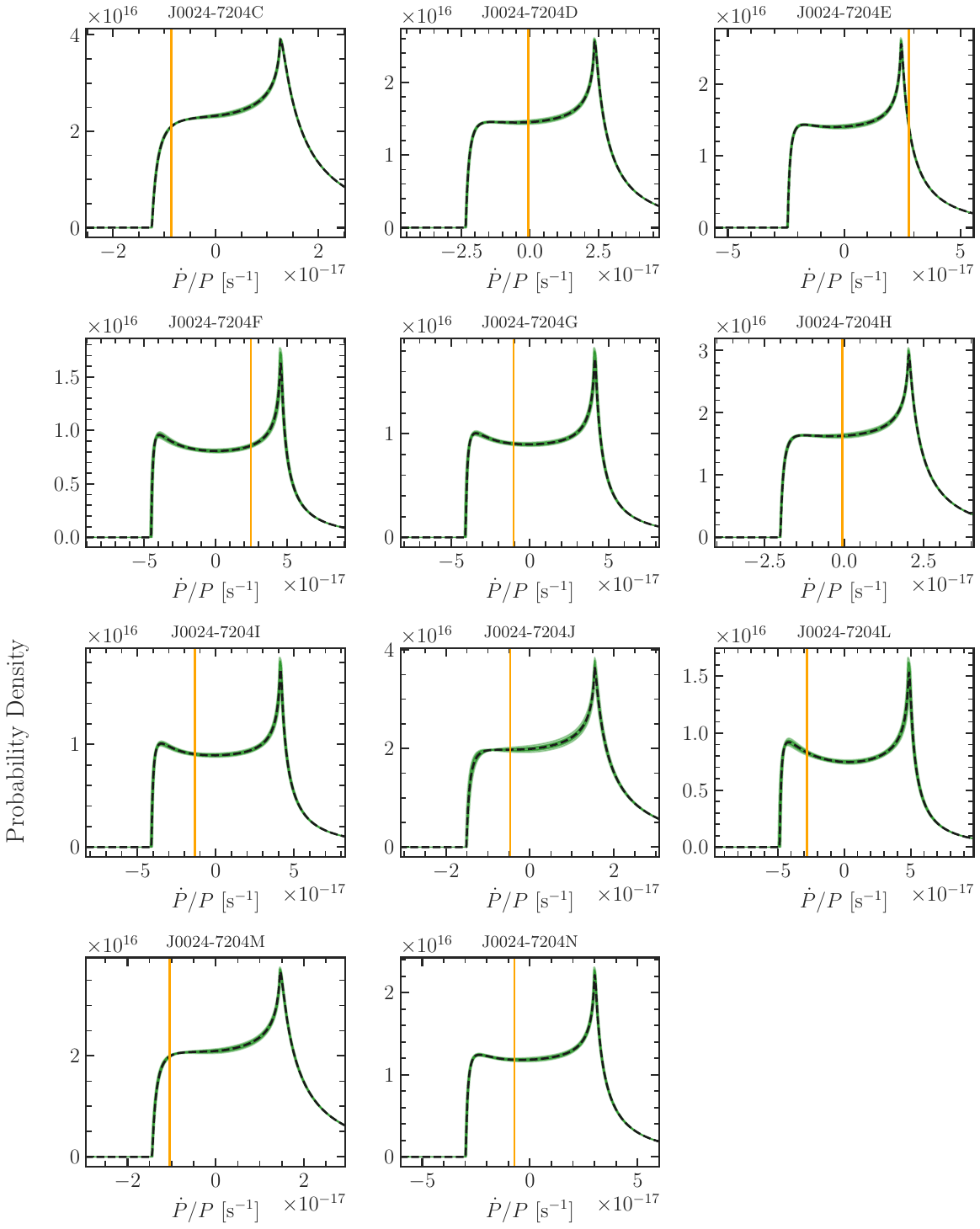}
	\caption{Likelihood functions corresponding to the best-fitting model (\TucAllData) for the observed \PdotP for each pulsar in 47 Tuc. In each panel, we show the observed period derivative as a vertical orange line.
	\label{fig:47Tuc-Paz-spin-grid1}}
\end{figure*}

\begin{figure*}
\centering
	\includegraphics[width=0.8\textwidth]{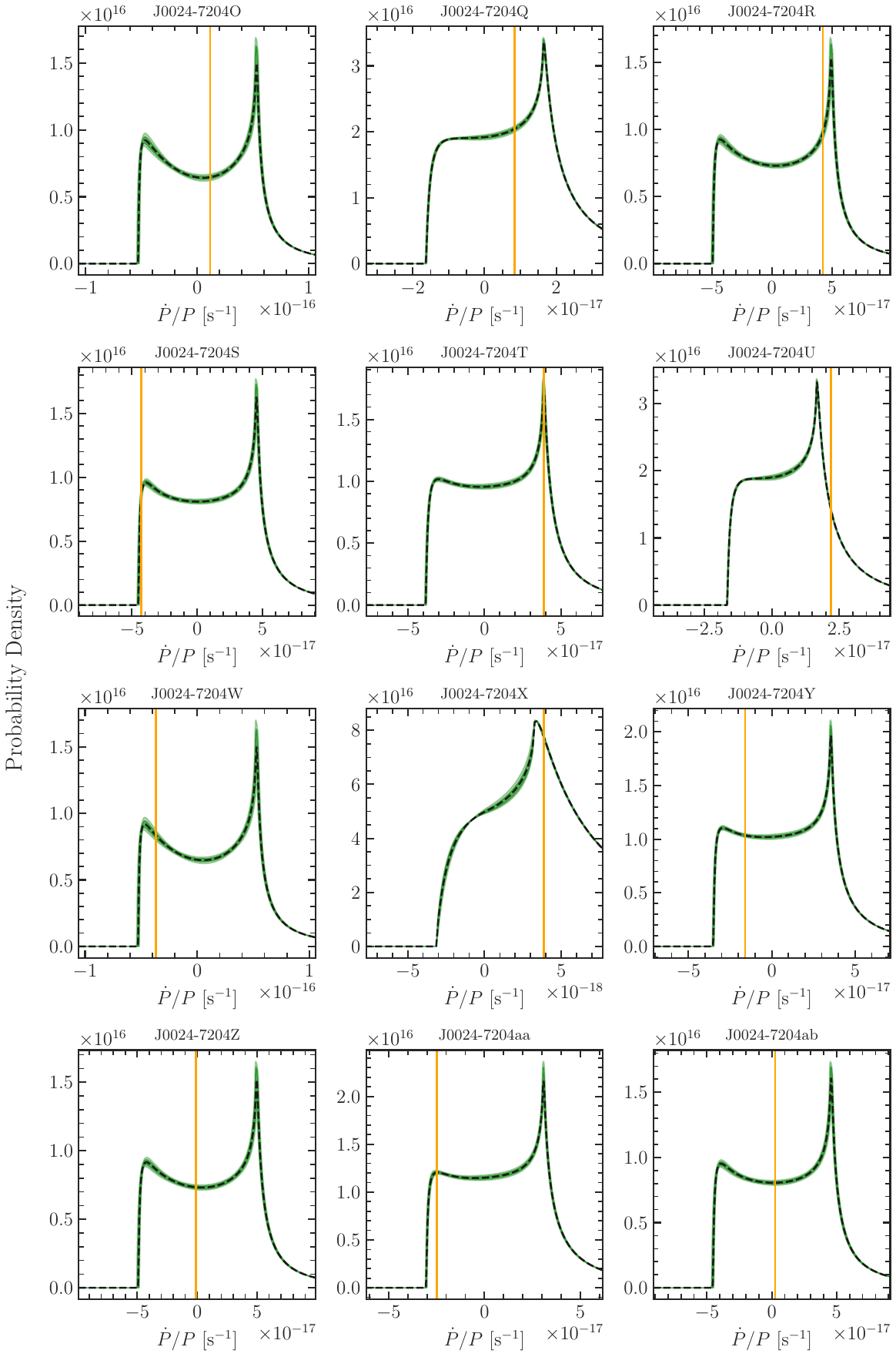}
	\caption{Continuation of Figure \ref{fig:47Tuc-Paz-spin-grid1}.
	\label{fig:47Tuc-Paz-spin-grid2}}
\end{figure*}

\begin{figure*}
\centering
	\includegraphics[width=0.8\textwidth]{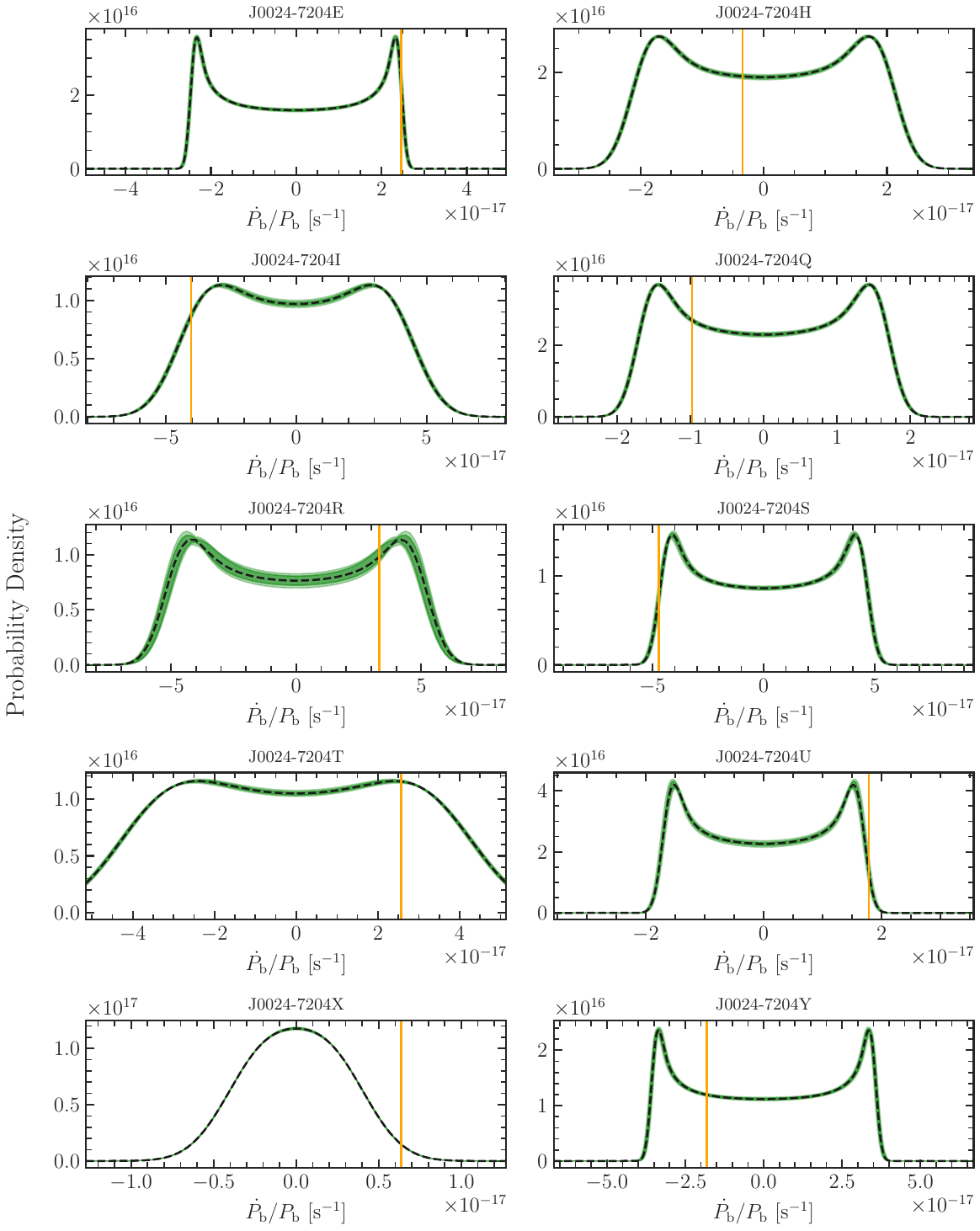}
	\caption{Likelihood functions corresponding to the best-fitting model (\TucAllData) for the observed \PbdotPb for each pulsar in 47 Tuc with an orbital timing solution. In each panel, we show the observed period derivative as a vertical orange line.
	\label{fig:47Tuc-Paz-orbital-grid}}
\end{figure*}

\begin{figure*}
\centering
	\includegraphics[width=0.8\textwidth]{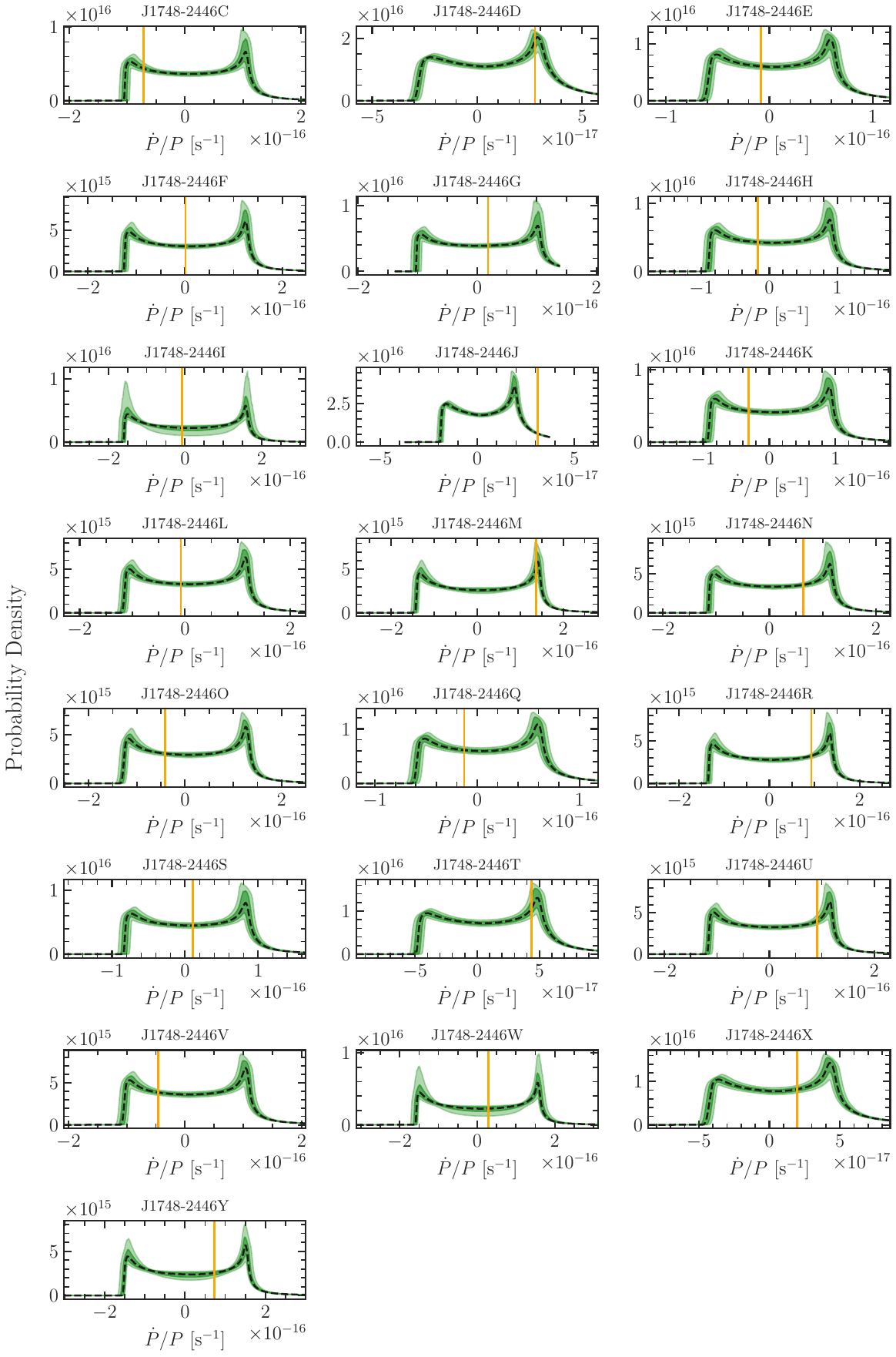}
	\caption{Likelihood functions corresponding to the best-fitting model (\TerAllData) for the observed \PdotP for each pulsar in Terzan 5. In each panel, we show the observed period derivative as a vertical orange line.
	\label{fig:Ter5-Paz-spin-grid1}}
\end{figure*}

\begin{figure*}
\centering
	\includegraphics[width=0.8\textwidth]{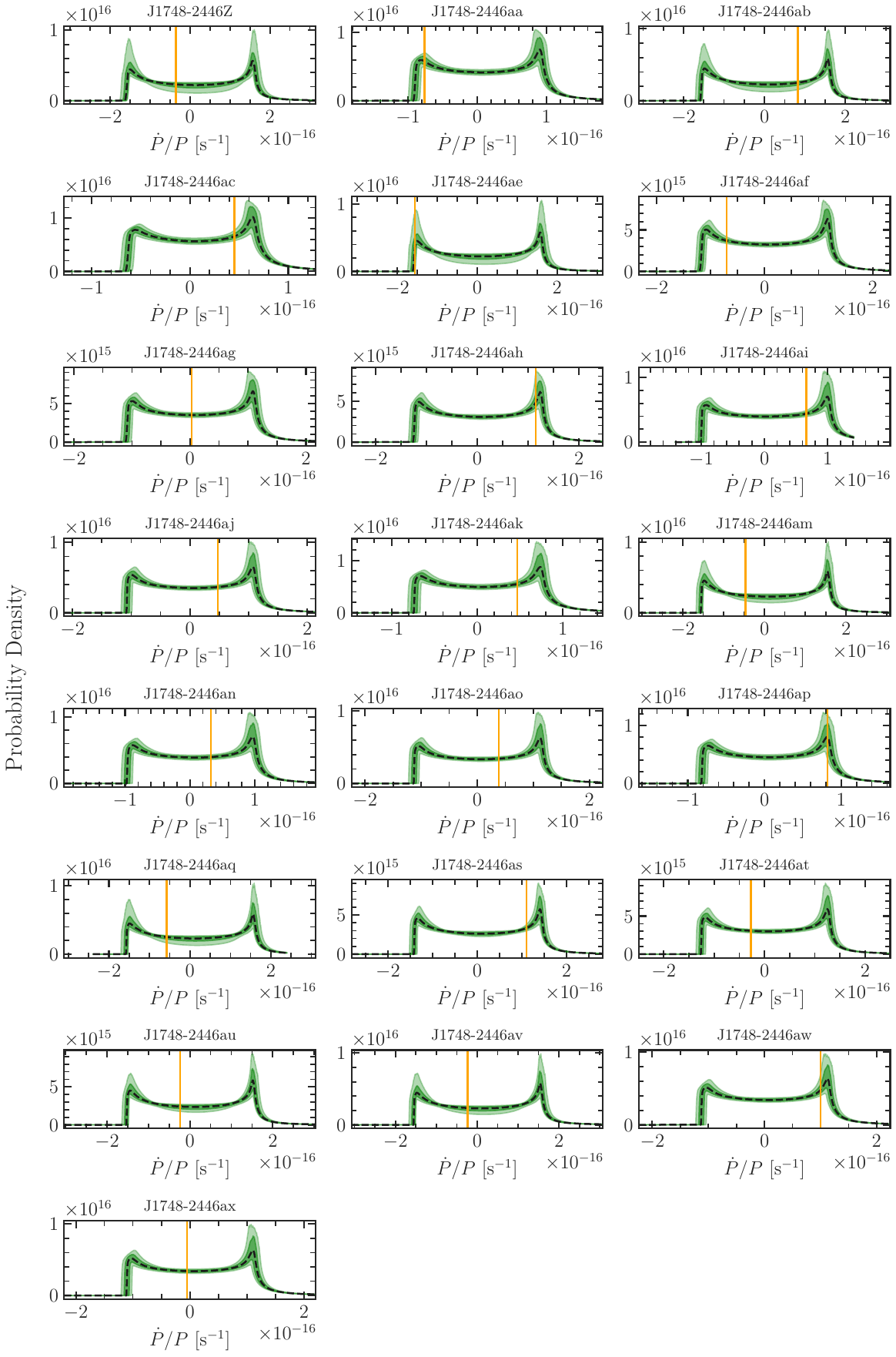}
	\caption{Continuation of Figure \ref{fig:Ter5-Paz-spin-grid1}.
	\label{fig:Ter5-Paz-spin-grid2}}
\end{figure*}

\begin{figure*}
\centering
	\includegraphics[width=0.8\textwidth]{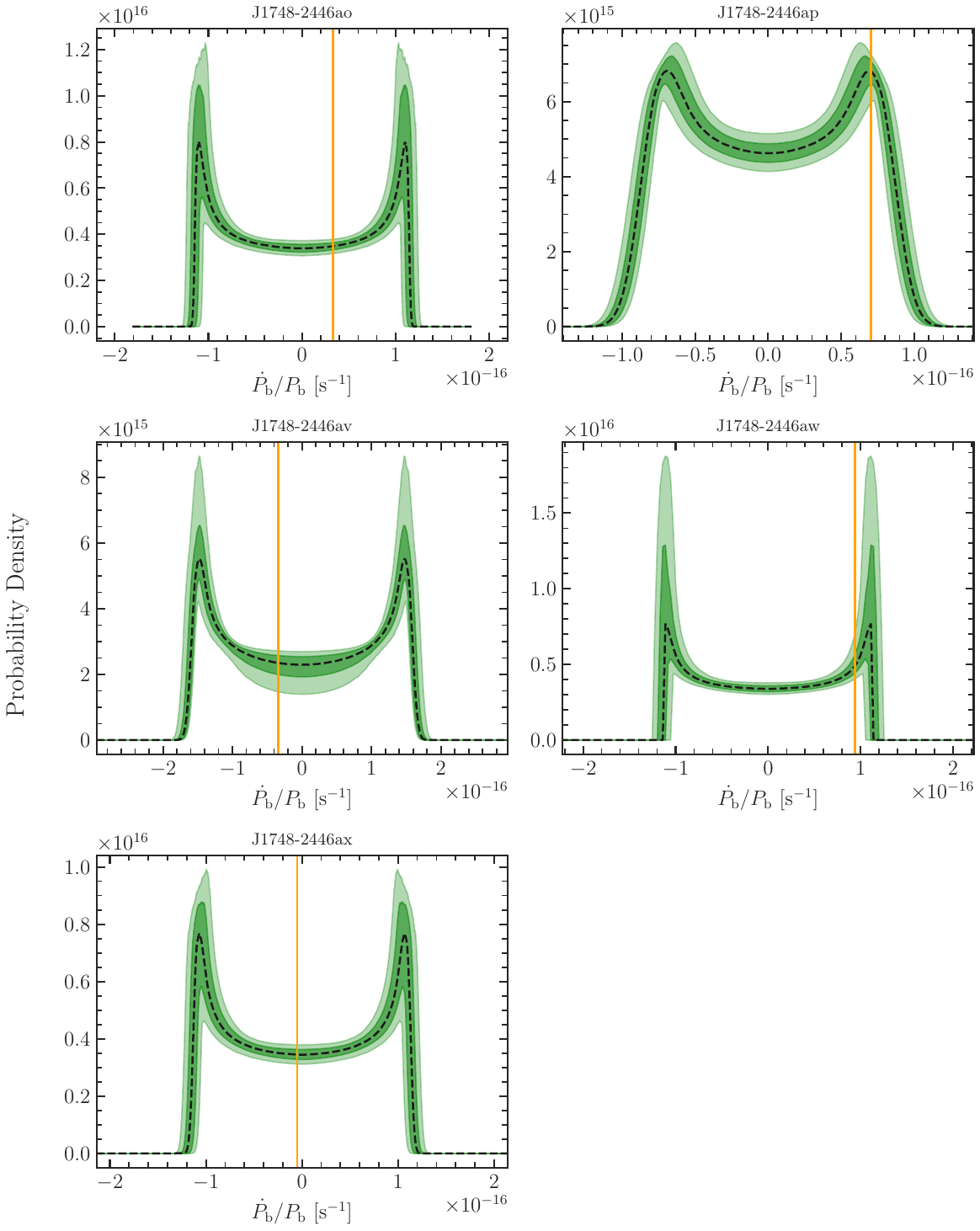}
	\caption{Likelihood functions corresponding to the best-fitting model (\TerAllData) for the observed \PbdotPb for each pulsar in Terzan 5 with an orbital timing solution. In each panel, we show the observed period derivative as a vertical orange line.
	\label{fig:Ter5-Paz-orbital-grid}}
\end{figure*}

%%%%%%%%%%%%%%%%%%%%%%%%%%%%%%%%%%%%%%%%%%%%%%%%%%


\bibliography{ms}{}
\bibliographystyle{aasjournal}



\end{document}